\documentclass[aps,prc,twocolumn,showpacs,floatfix,superscriptaddress,amsmath,amssymb,nofootinbib,longbibliography]{revtex4-1}
   
\usepackage{amsmath}
\usepackage{hyperref}
\usepackage{color}
\usepackage{graphicx}

\newcommand{\be}{\begin{equation}}
\newcommand{\ee}{  \end{equation}}
\newcommand{\ba}{\begin{eqnarray}}
\newcommand{\ea}{  \end{eqnarray}}
\newcommand{\bas}{\begin{eqnarray*}}
\newcommand{\eas}{  \end{eqnarray*}}

\begin{document}

\title{Effective field theory for deformed odd-mass nuclei}

\author{T. Papenbrock} 

\affiliation{Department of Physics and Astronomy, University of
  Tennessee, Knoxville, Tennessee 37996, USA}
\affiliation{Physics Division, Oak Ridge National Laboratory, Oak Ridge, Tennessee 37831, USA}

\author{H. A. Weidenm{\"u}ller}

\affiliation{Max-Planck-Institut f{\"u}r Kernphysik, D-69029
  Heidelberg, Germany}

\begin{abstract}
We develop an effective field theory (EFT) for deformed odd-mass
nuclei. These are described as an axially symmetric core to which a
nucleon is coupled. In the coordinate system fixed to the core the
nucleon is subject to an axially symmetric potential. Power counting
is based on the separation of scales between low-lying rotations and
higher-lying states of the core. In leading order, core and nucleon
are coupled by universal derivative terms. These comprise a covariant
derivative and gauge potentials which account for Coriolis forces and
relate to Berry-phase phenomena.  At leading order, the EFT combines
the particle-rotor and Nilsson models. We work out the EFT up to
next-to-leading order and illustrate the results in $^{239}$Pu and
$^{187}$Os. At leading order, odd-mass nuclei with rotational band
heads that are close in energy and differ by one unit of angular
momentum are triaxially deformed. For band heads that are well
separated in energy, triaxiality becomes a subleading effect. The EFT
developed in this paper presents a model-independent approach to the
particle-rotor system that is capable of systematic improvement.
\end{abstract}

\maketitle
\section{Introduction}
\label{intro}
In the last two decades, ideas based on effective field theory (EFT)
and on the renormalization group have exerted a strong influence on
nuclear-structure
theory~\cite{bedaque2002,bogner2003,epelbaum2009,bogner2010,griesshammer2012,hammer2017}.
These ideas have led to model-independent approaches to nuclear
interactions, currents, and nuclear spectra, to a new understanding of
resolution-scale and scheme dependences in theoretical
calculations~\cite{bedaque1999,furnstahl2002b,anderson2010}, and to
quantitative estimates of theoretical
uncertainties~\cite{schindler2009,furnstahl2014c}. EFT exploits a
separation of scale between the low-energy phenomena of interest and
the excluded high-energy aspects. Thus, EFT can also be used to
describe low-lying collective nuclear excitations such as
rotations~\cite{papenbrock2011,papenbrock2014,papenbrock2015,papenbrock2016,chen2017,chen2018,chen2020}
and vibrations~\cite{coelloperez2015b,coelloperez2016}. Venerable
nuclear collective models~\cite{eisenberg1970,bohr1975,iachello}
have been identified as leading-order Hamiltonians in an EFT approach.

In this work, we develop an EFT for odd-mass deformed nuclei. These
are viewed as a nucleon coupled to an axially symmetric core. Many
even-even deformed nuclei exhibit some amount of triaxiality even in
low-lying rotational bands. That, however, is often a small effect
that can be treated as a higher-order correction to a first-order
description that uses axial symmetry. Our approach differs from the
general particle-rotor model and from a very recently developed
EFT~\cite{chen2020}, both of which couple the nucleon to a triaxially
deformed nucleus. As we will see below, the coupling of a nucleon to
an axially symmetric core can, however, yield a triaxially deformed
nucleus.

The theoretical arguments that lead to the Hamiltonian of the
particle-rotor model are deceptively
simple~\cite{herzberg1945,bohr1952,bohr1953,nilsson1955,kerman1956}:
In the body-fixed (i.e. co-rotating) coordinate system (indicated here
and in what follows by primes), a particle with angular momentum
$\mathbf{K}=(K_{x'},K_{y'},K_{z'})$ is coupled to a rotor with angular
momentum $\mathbf{R}=(R_{x'},R_{y'},R_{z'})$, resulting in the total
angular momentum $\mathbf{I} =\mathbf{R}+\mathbf{K}$. The Hamiltonian
of a rotor is given by
\ba
\label{Hpr}
H &=& \sum_{k = x',y',z'} {R_k^2\over 2C_k} \ . 
\ea
Here, $C_k$ are the moments of inertia. Replacing the components of
$\mathbf{R}$ by those of $\mathbf{I} - \mathbf{K}$ leads to the
Hamiltonian
\ba
\label{Hpr1}
H_{\rm rot} = \sum_{k = x',y',z'} {(I_k - K_k)^2\over 2C_k}
\ea
of the particle-rotor model. That model describes a wealth
of data on odd-mass nuclei.

We are motivated to develop an EFT for the particle-rotor model
because that approach is expected to yield a systematic classification
of terms in the Hamiltonian according to their order of importance,
with the Hamiltonian~(\ref{Hpr1}) expected to appear as the
leading-order term. For that, the formulation of the particle-rotor
model in terms of angular momenta is not a good starting point,
however. In line with common usage, our EFT is based upon the
Lagrangian or the Hamiltonian formalism. These, in turn, make use of
velocities or canonical momenta, respectively. However, a Lagrangian
approach to the particle-rotor system is not contained in the standard
textbooks~\cite{rowe1970,eisenberg1970,bohr1975,ringschuck,iachello,rowe2010}.

In addition to providing a systematic procedure for generating
Hamiltonian terms of given order, the EFT approach yields surprises
and interesting results. For example, the coupling between the
particle and the rotor can naturally be described in terms of Abelian
and non-Abelian gauge potentials. Such potentials, and the Berry
phases~\cite{simon1983,berry1984} associated with them, enter in the
description of diatomic
molecules~\cite{wilczek1984,jackiw1986,bohm1992} and the quantum Hall
effect~\cite{estienne2011}. However, Berry phases have received less
attention in low-energy nuclear
physics~\cite{nikam1987,bulgac1990,klein1993,nazarewicz1994,hayashi1997,chandrasekharan2008}.

This paper is organized as follows. We identify the relevant
low-energy degrees of freedom in Sect.~\ref{sec:dof}. In
Sect.~\ref{sec:EFT} we systematically construct the EFT by presenting
the power-counting procedure and introducing the relevant interactions
at leading and at next-to-leading order. Hamiltonian and total angular
momentum are introduced, the Hamiltonian is diagonalized and spetra in
leading and subleading order are calculated in Section~\ref{lagr}. We
present applications of our results to $^{239}$Pu and the triaxially
deformed $^{187}$Os in Sec.~\ref{sec:apps}, and summarize our results
in Sect.~\ref{sec:sum}. Numerous appendices give the technical details
necessary for a self-contained description.

\section{Degrees of freedom and separation of scales}
\label{sec:dof}

\subsection{Even-even nucleus: rotating core}
\label{eerot}

\begin{figure}[bt]
\includegraphics[width=0.45\textwidth]{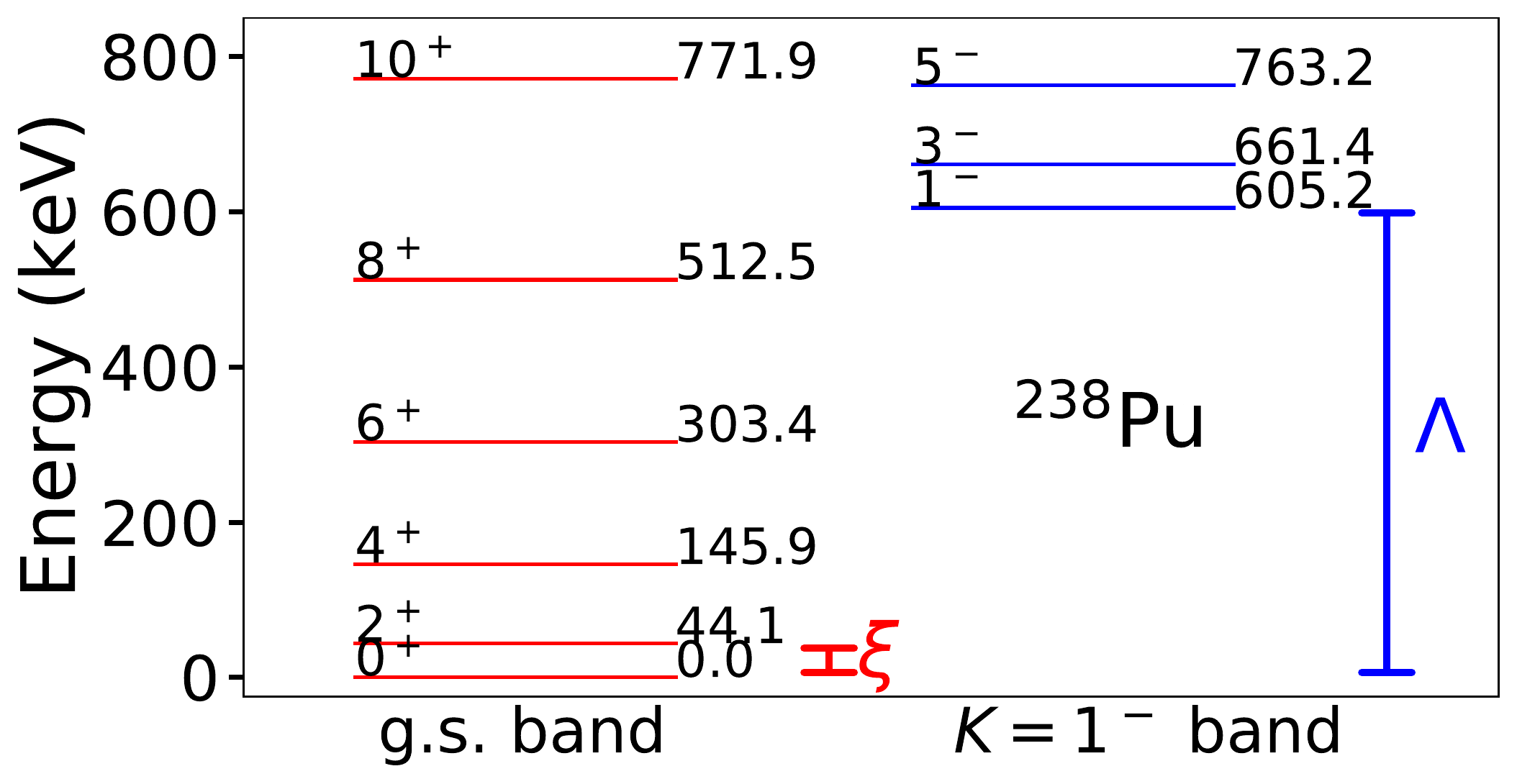}
\caption{(Color online) The levels of $^{238}$Pu below 800~keV can be
  grouped in two rotational bands, with spin/parity and energy for
  each level as indicated. The low-energy scale $\xi \approx 40$~keV
  sets the scale for rotations. The breakdown scale $\Lambda \approx
  600$~keV indicates a ``vibrational" state, i.e. the breakdown of the
  axially-symmetric rigid-rotor picture for this nucleus.  }
\label{fig:pu238}
\end{figure}

Many odd-mass deformed nuclei can be viewed as an even-even deformed
nucleus to which the extra nucleon is coupled. We take $^{239}$Pu as
an example. The corresponding even-even nucleus is $^{238}$Pu, and
Fig.~\ref{fig:pu238} shows all its levels below 800~keV. At
sufficiently low energies the spectrum of $^{238}$Pu is essentially
that of an axially symmetric rigid rotor: The excitation energies
$E(I)$ versus angular momentum $I$ obey $E(I) = AI(I+1)$. Here $A$ is
a rotational constant of about 7~keV, and $\xi \approx 40$~keV (the
energy of the $I=2$ state) sets the low-energy scale.  Only even spins
enter because the ground state is invariant under rotations by $\pi$
around any axis that is perpendicular to the symmetry axis. This
symmetry is usually denoted as $R$ symmetry~\cite{bohr1975}. At energy
$\Lambda \approx 600$~keV a second rotational band with a $K^\pi=1^-$
band head occurs, followed by more rotational bands at higher
energies. In this work we will, however, consider only the lowest
energies and restrict ourselves to the description of the ground-state
rotational band.  Then, the energy of the $K^\pi=1^-$ band head sets
the breakdown scale $\Lambda$ of our EFT, because a new degree of
freedom enters at this energy.  We have a separation of scale $\xi \ll
\Lambda$. The analysis of Ref.~\cite{coelloperez2015} shows that the
ground-state band will exhibit noticeable deviations from the
leading-order $E(I) = AI(I+1)$ rule for spins $I\gtrsim
\Omega/\xi$. In the EFT this is due to subleading interactions that
couple the ground-state band to other bands. While the interaction
between the positive-parity ground-state band and the shown
negative-parity band is suppressed, a positive-parity band enters at
about 940~keV.  These arguments suggest that the breakdown scale is
properly chosen.  An EFT for the lowest energies in deformed nuclei
was presented in Refs.~\cite{papenbrock2011,coelloperez2015}, and we
briefly review its essential features.

\subsubsection{The rotor in quantum mechanics}
Nuclear deformation causes an emergent breaking~\cite{yannouleas2007}
of rotational symmetry from SO(3) to axial SO(2), described as a
nonlinear realization of the
symmetry~\cite{weinberg1968,coleman1969,callan1969,leutwyler1994,weinberg_v2_1996,brauner2010}. The
degrees of freedom corresponding to the remnants of Nambu-Goldstone
bosons parametrize the coset SO(3)/SO(2), i.e. the two-sphere. We use
the radial unit vector
\ba
\label{coordinates}
\mathbf{e}_r &\equiv& \cos\phi \sin\theta \mathbf{e}_{x} + \sin\phi
\sin\theta \mathbf{e}_y +\cos\theta   \mathbf{e}_z 
\ea
for this purpose. Here, $(\mathbf{e}_{x}, \mathbf{e}_{y},
\mathbf{e}_{z})$ are orthogonal unit vectors that span a right-handed
coordinate system (the ``space-fixed system''), and $\theta$ and
$\phi$ are the polar and azimuthal angle, respectively.  The vector
$\mathbf{e}_r$ in Eq.~(\ref{coordinates}) points in the direction of
the symmetry axis of the deformed nucleus. It is supplemented by the
unit vectors
\ba
\mathbf{e}_{\theta} &\equiv& \cos\phi \cos\theta \mathbf{e}_x +\sin\phi
\cos\theta \mathbf{e}_y -\sin\theta \mathbf{e}_z \ , \nonumber\\
\mathbf{e}_{\phi} &\equiv& 
-\sin\phi \mathbf{e}_x +\cos\phi \mathbf{e}_y \ .
\ea
The vectors $(\mathbf{e}_\theta, \mathbf{e}_\phi, \mathbf{e}_r)$ span
the (right-handed) ``body-fixed'' coordinate system of the rotor. They
result from rotating the axes $\mathbf{e}_{x}$, $\mathbf{e}_{y}$, and
$\mathbf{e}_{z}$ of the space-fixed system by the operator ${\cal
  R}(\phi, \theta, 0)$. Here ${\cal R}$ stands for the general
rotation
\be
\label{rot}
   {\cal R}(\alpha,\beta,\gamma)\equiv e^{-i\alpha J_z} e^{-i\beta J_y}
   e^{-i\gamma J_z} \ , 
\ee
parametrized in terms of the Euler angles $(\alpha,\beta,\gamma)$. The
operators $J_k$ with $k = x, y, z$ generate rotations around the axes
$\mathbf{e}_k$ and fulfill the usual commutation relations
\be
[J_x, J_y] = i J_z \ {\rm (cyclic)} \ .
\label{comm}
\ee
We also use the notation
\ba
\label{eprime}
\mathbf{e}^\prime_{x} &=& \mathbf{e}_\theta \ , \nonumber\\
\mathbf{e}^\prime_{y} &=& \mathbf{e}_\phi \ , \nonumber\\
\mathbf{e}^\prime_{z} &=& \mathbf{e}_r 
\ea
for the basis vectors of the body-fixed coordinate system.

In addition to the generators $(J_x, J_y, J_z)$ of rotations in the
space-fixed system we also use their analogues $(J_{x'}, J_{y'},
J_{z'})$ in the body-fixed system. These also obey the commutation
relations~(\ref{comm}). If space-fixed and body-fixed system
originally coincide, the rotation~(\ref{rot}) and the rotation
\be
\label{rot'}
   {\cal R}^\prime(\alpha,\beta,\gamma) \equiv e^{-i\gamma J_{z'}} e^{-i\beta J_{y'}}
   e^{-i\alpha J_{z'}}
\ee
are identical~\cite{varshalovich1988}. For $\alpha = \phi$, $\beta =
\theta$ the last two factors in expression~(\ref{rot'}) rotate the
space-fixed $z$-axis into the direction of ${\bf e}'_z$. The remaining
factor $e^{-i\gamma J_{z'}}$ rotates the resulting system about the
body-fixed ${\bf e}'_z$-axis. Hence, an operator defined in the
body-fixed system that is invariant under SO(2) rotations, is
automatically invariant under general SO(3) rotations in the
space-fixed system. We use that insight to construct invariant terms
in the Lagrangian.
  
Our definition~(\ref{eprime}) of the body-fixed coordinate system,
resulting from the application of the rotation ${\cal
  R}(\phi,\theta,0)$ to the space-fixed system, represents but one
possibility. Any rotation ${\cal R}(\phi,\theta,\gamma)$ with
$\gamma=\gamma(\theta,\phi)$ of the space-fixed system would be
equally acceptable (albeit $\gamma= 0$ seems particularly simple).  As
we will see below, this arbitrary convention leads to a gauge
freedom~\cite{littlejohn1997}.

The time-dependent angles $(\theta, \phi)$ describe the motion of the
deformed nucleus. The angular velocity is
\ba
\label{velocity}
\mathbf{v} &\equiv& \frac{d}{dt} \mathbf{e}_r \nonumber\\
&=& v_\theta \mathbf{e}_\theta +v_\phi \mathbf{e}_\phi \ ,
\ea
with
\ba
\label{v_theta_phi}
v_\theta &\equiv & \dot{\theta}  \ , \nonumber\\
v_\phi &\equiv&  \dot{\phi}\sin\theta \ .
\ea
The dot denotes the time derivative. We see that the rotor's degrees
of freedom transform non-linearly [i.e. they depend in a nonlinear way
  on $(\phi, \theta)$] under the rotation. The expression
$\mathbf{v}^2$ with $\mathbf{v}$ defined in Eq.~(\ref{velocity}) is
obviously invariant and so is, therefore, the Lagrangian
\ba
\label{Lrot}
L_{\rm rot} &=& {C_0\over 2} \mathbf{v}^2 = {C_0\over 2} \left(
\dot{\theta}^2 +\dot{\phi}^2 \sin^2\theta \right) \ .
\ea
This is, of course, the Lagrangian of an axially symmetric rotor (or,
equivalently, that of a particle on the unit sphere). Here $C_0$ is a
low-energy constant and corresponds to the moment of inertia.

We introduce the canonical momenta $p_\theta = {\partial L_{\rm
    rot} / \partial \dot{\theta}}$ and $p_\phi = {\partial L_{\rm
    rot} / \partial \dot{\phi}}$ and perform a Legendre transform of
the Lagrangian~(\ref{Lrot}). This yields the Hamiltonian
\ba
\label{Hrot}
H_{\rm rot} &=& {1\over 2C_0}\left(p_\theta^2 +{p_\phi^2\over\sin^2\theta}
\right) = {\mathbf{p}^2\over 2C_0} \ . 
\ea
Here, we combined the canonical momenta into
\be
\mathbf{p} \equiv p_\theta \mathbf{e}_\theta +{p_\phi\over \sin\theta}
\mathbf{e}_\phi \ .
\ee
We quantize the momentum $\mathbf{p}$ as usual,
\ba
\label{quant_mom}
\mathbf{p}=-i\nabla_\Omega \ , 
\ea
with 
\be
\nabla_\Omega \equiv \mathbf{e}_\theta\partial_\theta +{\mathbf{e}_\phi
  \over\sin\theta}\partial_\phi \ .
\ee
The spectrum is 
\be
\label{Erot}
E(I)={I(I+1)\over 2C_0} \ ,
\ee
with angular momenta $I$, corresponding to a rotational band.

An alternative derivation of the rotor spectrum uses the
  angular momentum
\be
\mathbf{I}=\mathbf{e}_r\times\mathbf{p} \ , 
\ee
rewrites, $\mathbf{p}=-\mathbf{e}_r\times\mathbf{I}$ (which implies
$\mathbf{p}^2=\mathbf{I}^2$), and thereby obtains the Hamilitonian
$\mathbf{I}^2/(2C_0)$.  We will use such an approach below.

\subsubsection{Connection to effective field theory}

The arguments in Section II.A.1 may seem purely phenomenological. We
now establish the connection to EFT. For nonrelativistic quantum
systems, that approach is summarized in Ref.~\cite{brauner2010}, see
also Ref.~\cite{leutwyler1994}. A paradigmatic application is that to
the infinitely extended ferromagnet (FM)
(Refs.~\cite{roman1999,hofmann1999,baer2004,kampfer2005}). The
breaking of a symmetry of the Hamiltonian in the ground state of the
system (in the FM: the common direction of all spins violates
rotational invariance) causes the existence of Nambu-Goldstone modes
(in the FM: spin waves). These make up for the fact that the FM cannot
rotate. They determine the low-lying part of the spectrum of the FM,
are determined entirely by the broken symmetry, and depend upon a
small number of parameters that must be fitted to the data. In atomic
nuclei, that EFT scheme must be generalized as we deal with ``emergent
symmetry breaking'', see
Refs.~\cite{yannouleas2007,papenbrock2014,papenbrock2015}. In the
limit of infinte mass, nuclei cannot rotate either. The
Nambu-Goldstone modes are surface vibrations. Finite nuclei are able
to undergo rotations, however. The associated degrees of freedom are
the purely time-dependent angles $\theta(t)$ and $\phi(t)$ introduced
in Section~II.A.1. These degrees of freedom are not Nambu-Goldstone
modes as they cease to carry physical significance in the
infinite-mass limit. Rather they represent the onset of symmetry
breaking in the finite system (hence ``emergent symmetry
breaking''). That approach is expected to work for systems close to
full symmetry breaking. Then the relevant energy scales are (in
increasing order) the rotational energy (via the degrees of freedom
$\theta$ and $\phi$), the surface vibrations (described in terms of
Nambu-Goldstone modes), and genuine intrinsic excitations of the
system (not accessible in terms of Nambu-Goldstone modes). In
Refs.~\cite{papenbrock2014,papenbrock2015} that approach has been
worked out in detail for even-mass nuclei. In the present paper we
confine ourselves in the description of the core to the very lowest
part of the excitation spectrum, i.e., to rotations.

Needless to say we may re-formulate the quantum mechanics of the
axially symmetric rotor as a quantum field theory.  Based on the
familiar rotor Hamiltonian~(\ref{Hrot}) we introduce the quantum field
operator $\hat{\Phi}(\theta,\phi)$ that fulfills the canonical
commutation relations for bosons
\be
\left[ \hat{\Phi}(\theta,\phi), \hat{\Phi}^\dagger(\theta',\phi')\right] = \delta(\phi-\phi')\delta(\cos\theta-\cos\theta') \ .
\ee
Here, $\hat{\Phi}^\dagger(\theta,\phi)$ creates an axially symmetric rotor
whose symmetry axis points into the direction of
$\mathbf{e}_r$.

The Lagrangian of the free rotor is then
\begin{align}
  \label{QFT-L}
  L = \int\limits_0^{2\pi} {\rm d}\phi \int\limits_{-1}^1{\rm d}\cos\theta\,\,
  \hat{\Phi}^\dagger(\theta,\phi)\left(i\partial_t + {\nabla_\Omega^2\over 2C_0}\right)\hat{\Phi}(\theta,\phi) \ . 
\end{align}
Introducing the momentum operator
\be
\hat{\Pi}_\Phi(\theta,\phi)\equiv {\delta L\over \delta\partial_t\hat{\Phi}(\theta,\phi)}
= i\hat{\Phi}^\dagger(\theta,\phi)
\ee
and performing the usual Legendre transformation then yields the Hamiltonian
\begin{align}
  H = - {1\over 2C_0}\int\limits_0^{2\pi} {\rm d}\phi \int\limits_{-1}^1{\rm d}\cos\theta \,\,
  \hat{\Phi}^\dagger(\theta,\phi)\nabla_\Omega^2\hat{\Phi}(\theta,\phi) \ . 
\end{align}
This clearly is the second-quantized version of the
Hamiltonian~(\ref{Hrot}).

Although we are dealing with the rotor in quantum mechanics and not in
quantum field theory we continue to use the terminology of EFT.  This
is in keeping with many works in low-energy nuclear physics where the
ideas of EFTs~\cite{lepage1997} are used to construct and solve
Hamiltonians in quantum mechanics, see, e.g.,
Refs.~\cite{scaldeferri1997,beane1998,kirscher2010,lensky2016,capel2018}.

\subsection{Nucleon}
\label{sec:fermion}

\begin{figure*}
\includegraphics[width=0.96\textwidth]{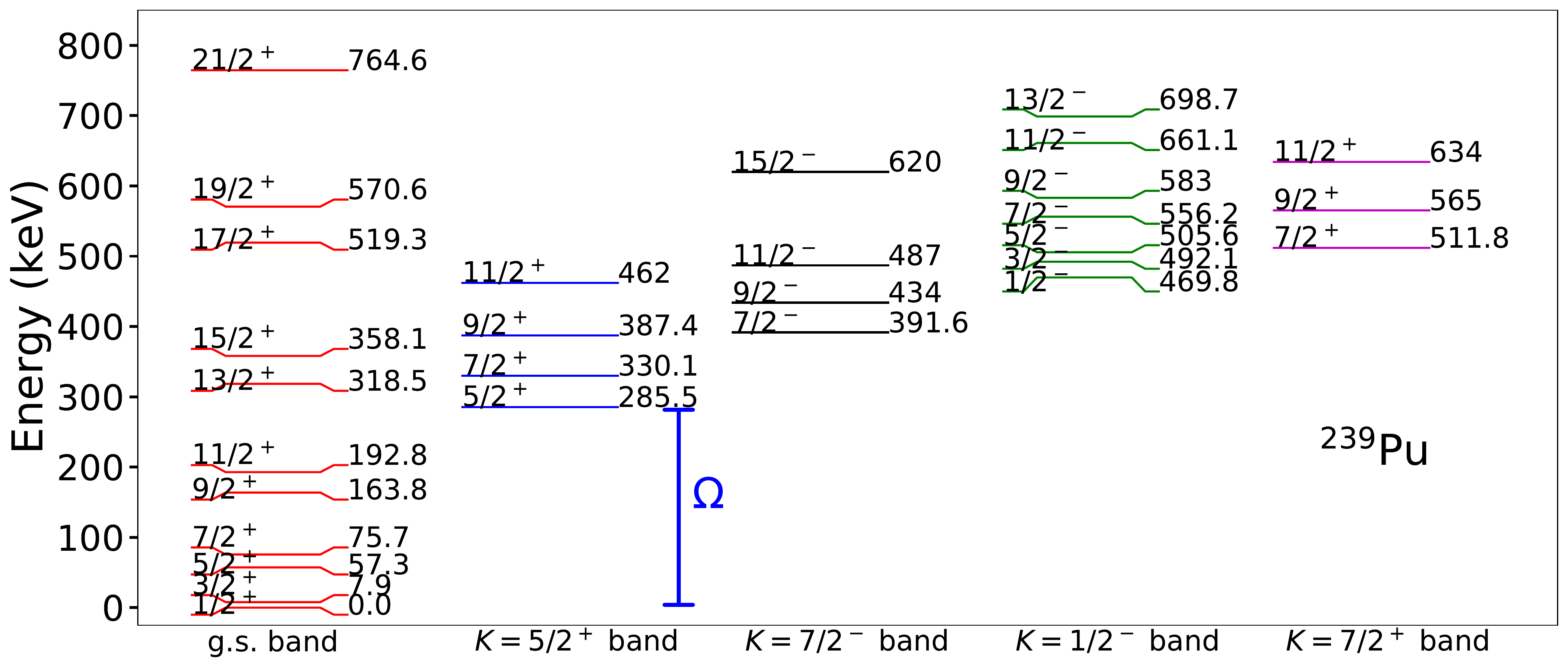}
\caption{(Color online) Levels of $^{239}$Pu below 800~keV that can be
  grouped into rotational bands, with spins, parities, and energies as
  indicated. The energy $\Omega \approx 300$~keV sets the scale for
  single-particle excitations.  }
\label{fig:pu239}
\end{figure*}

To gain insight into how to construct the EFT, we consider the
odd-mass nucleus $^{239}$Pu. Figure~\ref{fig:pu239} shows all levels
below 800~keV that can be grouped into rotational bands (omitting the
few exceptions). The ground-state rotational band is built on a
$K^\pi={1\over 2}^+$ state, i.e., a $K^\pi={1\over 2}^+$ neutron
coupled to the $^{238}$Pu ground state. Rotations of this
nucleon-nucleus state then produce the rotational band on top of the
$1/2^+$ ground state. The first excited neutron state yields the
$K^\pi={5\over 2}^+$ state at $\Omega\approx 300$~keV, and its
rotations produce the corresponding rotational bands. Thus, the
fermion single-particle excitation energy is about half the breakdown
scale in this nucleus, and the condition $\Omega \ll \Lambda$ is
fulfilled only marginally. The $K^\pi={1\over 2}^-$ band head at about
470~keV could be due either to a single-neutron excitation or to the
coupling of the $K^\pi={1\over 2}^+$ neutron with the excited $1^-$
state (at the breakdown energy $\Lambda$) in $^{238}$Pu. Therefore,
that rotational band is beyond the breakdown scale of the EFT we
present in this paper.

The rotational bands depicted in Fig.~\ref{fig:pu239} all follow the
pattern
\be
\label{spectrum}
E(I,K) = E_0 + A \left[ I(I+1) +a \delta_{K, 1/2} (-1)^{I+{1\over 2}}
  \left( I+{1\over 2} \right) \right] \ .
\ee
Here, $E_0$ is an energy offset, $A$ the rotational constant, and $a$
the decoupling parameter (that occurs only for $K={1\over 2}$
bands). These constants depend on the band under
consideration. Typically, we have $E_0\sim\Omega$, $A\sim\xi/6$, and
$a\sim{\cal O}(1)$. Equation~(\ref{spectrum}) is well known from a
variety of
models~\cite{rowe1970,eisenberg,bohr1975,iachello,rowe2010}. As shown
below, it is also the leading-order result of the EFT we develop in
this paper.

We use the insight gained in the previous Subsection and request that
the Lagrangian of the nucleon be invariant under SO(2) rotations in
the body-fixed system. That guarantees invariance under SO(3)
rotations in the space-fixed system.

The field operator $\hat{\psi}_{s}(\mathbf{x}')$ creates a fermion at
position $\mathbf{x}'$ with spin projection $s = \pm {1\over2}$ onto
the $z'$-axis in the body-fixed frame.  Denoting the vacuum as
$|0\rangle$ we thus have
\be
\hat{\psi}_s^\dagger(\mathbf{x}')|0\rangle = \chi_{{1\over 2}s}|
\mathbf{x}'\rangle \ .
\ee
Here $\chi_{{1\over 2}s}$ denotes a spin state of spin-${1\over 2}$
fermion with $z'$ projection $s$~\cite{varshalovich1988}, and
$|\mathbf{x}'\rangle$ is an eigenstate of the position operator. The
corresponding annihilation operator is $\hat{\psi}_s(\mathbf{x}')$ and
we have the usual anti-commutation relation for fermions
\ba
\left\{\hat{\psi}_s(\mathbf{x}'), \hat{\psi}^\dagger_\sigma(\mathbf{y}')
\right\} = \delta_\sigma^s \delta(\mathbf{x}'-\mathbf{y}') \ , 
\ea
and all other anti commutators vanish. It will be useful to combine
the two spin components of the field operator into the spinor
\ba
\label{spinor-s}
\hat{\Psi}(\mathbf{x}') \equiv \left(\begin{array}{c}
\hat{\psi}_{+{1\over 2}}(\mathbf{x}')\\
\hat{\psi}_{-{1\over 2}}(\mathbf{x}')\end{array}\right) \ .
\ea
The nucleon Lagrangian is
\ba
\label{Lpsi}
L_{\Psi} = \int{\rm d}^3\mathbf{x}' \hat{\Psi}^\dagger(\mathbf{x}')
\left( i\partial_t + {\hbar^2\Delta' \over 2m}-V\right)
\hat{\Psi}(\mathbf{x}') \ .
\ea
Here, $V$ is an axially symmetric potential which may also depend on
spin, i.e., be a $2\times 2$ matrix. The potential of the Nilsson
model~\cite{nilsson1955} is an example. The Lagrangian~(\ref{Lpsi})
exhibits axial symmetry. The construction~(\ref{Lpsi}) is not only
mandated by the nonlinear realization of rotational
symmetry~\cite{papenbrock2011}. It is also consistent with an
adiabatic approach where the light nucleon is much faster than the
heavy and slowly rotating core and able to follow the core's motion
quasi instantaneously. The canonical momentum is
\be
\hat{\Pi}(\mathbf{x}') = \frac{\delta L}{\delta \partial_t
  \hat{\Psi}(\mathbf{x}')} = i \hat{\Psi}^\dag(\mathbf{x}') \ .
\ee
The Legendre transform of the Lagrangian~(\ref{Lpsi}) yields the
Hamiltonian
\ba
\label{Hpsi}
H_{\Psi} &=&  \int{\rm d}^3\mathbf{x}' \hat{\Psi}^\dagger(\mathbf{x}')
\left(  - {\hbar^2\Delta' \over 2m}+V\right) \hat{\Psi}(\mathbf{x}') \ .
\ea 
Here, we introduced the Laplacian $\Delta'=\nabla^\prime\cdot
\nabla^\prime$. The total angular momentum of the fermion,
\be
\label{Jferm}
\mathbf{K} = \int {\rm d}^3\mathbf{x}' \hat{\Psi}^\dagger
(\mathbf{x}')\left(-i \mathbf{x}'\times\mathbf{\nabla}' +
\hat{\mathbf{S}}\right) \hat{\Psi}(\mathbf{x}') \ , 
\ee
is the sum of orbital angular momentum and spin
\ba \hat{\mathbf{S}}={1\over 2}\left(\begin{array}{c} \sigma
  _{x'}\\ \sigma_{y'}\\ \sigma_{z'}\end{array}\right) \ .  \ea
Here, $\sigma_{x', y', z'}$ denote the usual Pauli matrices. These act
with respect to the axes of the body-fixed system. The action of the
general operators ($J_{x'}$, $J_{y'}$, $J_{z'})$ on the space- and
spin-degrees of freedom of the nucleon coincides with that of the
corresponding angular momentum plus spin operators in
Eq.~(\ref{Jferm}). Thus, for $k' = x', y', z'$,
\ba
K_{k'} \equiv \mathbf{e}^\prime_{k} \cdot \mathbf{K} =
\int {\rm d}^3\mathbf{x}' \hat{\Psi}^\dagger(\mathbf{x}') J_{k'}
\hat{\Psi}(\mathbf{x}') 
\label{nJ1}
\ea
are the projections of the fermion's angular momentum onto the
body-fixed axes.

Fermion states of axially-symmetric Hamiltonians $H_\Psi$ are written
as $|K, \alpha \rangle$. Here $K$ denotes the projection of the
angular momentum onto the nuclear symmetry axis, while $\alpha$
comprises the remaining quantum numbers energy, parity, and third
component of isospin. Kramers' degeneracy (i.e., time-reversal
invariance) implies that the single-Fermion states come in degenerate
pairs $|{\pm K},\alpha\rangle$, carrying identical quantum numbers
$\alpha$ and, in particular, sharing the same energy
$E_{|K|,\alpha}$. Thus, we have
\begin{align}
\label{Hferm2}
H_\Psi |K,\alpha \rangle &= E_{|K|,\alpha} |K, \alpha\rangle \ ,
\nonumber \\
\hat{K}_{z'} |K,\alpha \rangle &= K |K, \alpha\rangle \ .
\end{align}
We understand the band heads in Fig.~\ref{fig:pu239} simply as energy
eigenvalues of the fermion Hamiltonian $H_\Psi$ with a suitably chosen
potential $V$. The $R$ symmetry of the nuclear ground state ensures
that only a suitable linear combination of the states $|{\pm
  K},\alpha\rangle$ enters.  The energies $E_{|K|,\alpha}$ of the band
heads are of the scale $\Omega$. Thus, if we had spontaneous rather
than emergent symmetry breaking, $\xi\to 0$ would hold, and the
rotational bands on top of each band head would collapse.

The degrees of freedom of the rotor do not appear explicitly in
Eq.~(\ref{Hpsi}). Conversely, the potential $V$ has no impact on the
degrees of freedom of the rotor. The potential $V$ constitutes an
implicit interaction between the rotor and the particle which is
solely based on the fact that the potential is axially symmetric and
defined in the body-fixed frame. That is consistent with emergent
symmetry breaking which allows only a coupling to derivatives of
Nambu-Goldstone bosons, in our case: the angular velocity. Such
interactions -- not yet contained in the Hamiltonian~(\ref{Hpsi}) --
will appear as gauge couplings of the nucleon to the rotor. These are
partly constrained by the nonlinear realization of rotational
symmetry. They appear in universal form as a covariant derivative or
as Coriolis terms. They can also be understood within an adiabatic
approach that involves Berry phases.

\section{Building an effective field theory}
\label{sec:EFT}

Having identified in Section~\ref{sec:dof} the relevant degrees of
freedom due to a separation of scales, we now construct our effective
field theory for odd-mass deformed nuclei using the following
steps. (i) In the present Section we define the power-counting
procedure and (ii) write down the interaction terms between the
nucleon and the rotor in leading and some also in subleading order. In
step (ii) all possible interaction terms are admitted that are allowed
by the symmetries (in our case invariance under rotations, parity, and
time reversal), see, e.g., Ref.~\cite{brauner2010}. (iii) The resulting
Lagrangian and Hamiltonian, further subleading terms, and the solution
of the equations of motion are addressed in Section~\ref{lagr}.

\subsection{Power counting}

The power-counting procedure for the rotor was worked out in
Refs.~\cite{papenbrock2011,papenbrock2014}. We briefly present the
arguments. We associate the low-energy scale $\xi$ with the
rotor. Thus, the angular velocity scales as $\xi$,
\ba
|\mathbf{v}| &\sim& \xi \ , \nonumber\\
\dot{\theta} &\sim& \xi  \ , \nonumber\\
\dot{\phi} &\sim& \xi \ ,
\ea
and so does the Lagrangian~(\ref{Lrot}) of the free rotor.  That
implies that its low-energy constant scales as
\be
C_0\sim\xi^{-1} .
\ee

The spectrum of the free axially symmetric rotor forms a rotational
band, see Eq.~(\ref{Erot}), and $C_0$ is the moment of inertia. Let us
give examples. $C_0^{-1}\approx 1$~MeV for a light rotor such as
$^8$Be, 0.5~MeV in $^{24}$Mg, 0.2~MeV in the neutron-rich nucleus
$^{34}$Mg, 30~keV for a rare earth nucleus, and only 15~keV for
actinides.  These are the smallest energy scales in the nuclei we
consider. The breakdown energy $\Lambda$ for the rotor is set by
excitations that are not part of its ground-state rotational
band. This energy is about 17~MeV in $^8$Be, 4~MeV in $^{24}$Mg, 1~MeV
in rare earth nuclei, and about 0.5~MeV in actinides. Thus,
$\Lambda\gg\xi$ in all cases.

The subleading correction to the rotor Lagrangian~(\ref{Lrot}) is
\be
\label{C2}
C_2(\mathbf{v}^2)^2 \ . 
\ee
At the breakdown scale, i.e., when $|\mathbf{v}| \sim \Lambda$, the
term~(\ref{C2}) is by definition equal in importance to the
leading-order Lagrangian~(\ref{Lrot}). That yields
\be
C_2\sim\xi^{-1}\Lambda^{-2} \ .
\ee
At low energy where $|\mathbf{v}|\sim\xi$, the term~(\ref{C2}) yields
a contribution~$\sim \xi^3 / \Omega^2$ to the total Lagrangian, and
this is suppressed by $\xi^2/\Omega^2\ll 1$ compared to the leading
term~(\ref{Lrot}). That argument establishes the power-counting
procedure for the rotor: the energy scale $\xi$ is associated with
rotational bands. Corrections to the leading-order term come in powers
of $\xi/\Lambda$.

We turn to the energy scales of the Hamiltonian~(\ref{Hpsi}) of the
nucleon. The energy scale $\Omega$ is set by the mean level spacing of
the single-nucleon states, i.e., by the spacing of band-head energies
$E_{|K|\alpha}$ in odd-mass nuclei, see Eq.~(\ref{Hferm2}). That scale
is about 1.7~MeV in $^9$Be, (given by the energy difference of the
$3/2^-$ ground state and the excited $1/2^+$ band head), 0.6~MeV in
$^{25}$Mg, (given by the energy difference between the $5/2^+$ ground
state and the excited $1/2^+$ band head), and amounts to some hundreds
of keV in rare earth nuclei, and to tens to hundreds of keV for
actinides. In most odd-mass deformed nuclei we have $\Omega \gtrsim
\xi$, and in many cases one even finds $\Omega \gg \xi$, see
$^{239}$Pu in Fig.~\ref{fig:pu239} as an example. In such cases
$\Omega$ is only about a factor two or three away from the breakdown
scale $\Lambda$, and the separation between $\Omega$ and $\Lambda$ is
marginal so that $\Omega\lesssim\Lambda$ but not $\Omega\ll
\Lambda$. In such cases, the power counting uses both small expansion
parameters $\xi/\Lambda$ and $\xi/\Omega$, and it is difficult to
decide which is the more important one. If we had the ambition to
construct an EFT for the nucleon potential in the
Hamiltonian~(\ref{Hpsi}) we would have to deal, in addition, with an
expansion in powers of $\Omega/\Lambda$, but we do not pursue this
task in the present paper. In some nuclei such as $^{187}$Os discussed
below, we have $\Omega\sim \xi$ so that $\xi, \Omega \ll
\Lambda$. Then, the power counting is in $\xi/\Lambda$. [We avoid the
  equivalent parameter $\Omega/\Lambda$ as that might incorrectly
  suggest a systematic expansion of the nucleon potential in the
  Hamiltonian~(\ref{Hpsi}).]

In any case, the separation of scales allows us to construct an EFT
that systematicaly improves the energies and states in rotational
bands. We note that there are many states in odd-mass nuclei that do
not result from coupling a nucleon to the ground-state band of the
even-even nucleus (but rather from coupling to excited band heads of
the even-even nucleus). Such states fall outside the purview of the
EFT we aim to construct.  Including such effects would require us to
introduce fields that describe the non-rotational excitations of the
rotor.

It would be desirable to construct the potential $V$ in the fermion
Hamiltonian~(\ref{Hpsi}) in a similarly systematic fashion. We briefly
illuminate the difficulties in doing so for halo rotors, i.e.,
odd-mass nuclei where the nucleon is weakly bound to the even-even
core. Examples are $^9$Be (with a neutron separation energy of about
1.7~MeV) and neutron-rich magnesium isotopes with separation energies
below 1~MeV. In these cases, the fermion's de-Broglie wave length
exceeds the rotor's size, and a derivative expansion of the potential
seems appropriate. The potential $V$ must be axially symmetric. Total
spin $\hat{\mathbf{S}}^2=3/4$ and its projection $\hat{S}_{z'}^2=1/4$
are trivial constants, while $\hat{K}_{z'}$ is the nontrivial
conserved quantity and can be used to classify the fermion's wave
functions. Thus, we can parameterize the potential as
\ba
\label{Veft}
V &=& v_{01}\delta(\mathbf{r}') \nonumber\\
&+& v_{11}\nabla_\perp \delta(\mathbf{r}') \cdot\nabla_\perp +
v_{12}\partial_{z'} \delta(\mathbf{r}') \partial_{z'} \nonumber \\
&+&  v_{13}\left[\nabla^{2}_\perp\delta(\mathbf{r}') +
  \delta(\mathbf{r}') \nabla^{2}_\perp\right]\nonumber\\
&+&  v_{14}\left[\partial_{z'}^2\delta(\mathbf{r}') +
  \delta(\mathbf{r}')\partial_{z'}^2\right]\nonumber\\
&+& v_{15}\left(\hat{\sigma}_{x'}\partial_{x'} + \hat{\sigma}_{y'}
\partial_{y'}\right)\delta(\mathbf{r}')\left(\hat{\sigma}_{x'}
\partial_{x'}+\hat{\sigma}_{y'}\partial_{y'}\right) \nonumber\\
&+& v_{16}\big[\left(\hat{\sigma}_{x'}\partial_{x'} +
  \hat{\sigma}_{y'}\partial_{y'}\right)\hat{\sigma}_{z'}
  \partial_{z'}\delta(\mathbf{r}') \nonumber\\
  && \, \, \, + \, \, \, \, \delta(\mathbf{r}') \hat{\sigma}_{z'}
  \partial_{z'} \left(\hat{\sigma}_{x'}\partial_{x'} +
  \hat{\sigma}_{y'}\partial_{y'}\right) \big] \nonumber\\
&+& \cdots \ .
\ea
Here, $\nabla_\perp\equiv r^{-1}\nabla_\Omega$. In Eq.~(\ref{Veft}) we
did not present all second-order derivatives, and higher-order
derivatives are missing as well. If the fermion has quantum numbers
$J^\pi={1\over 2}^+$, the leading-order contribution consists solely
of the $v_{01}$ contact coupling. For $J^\pi={3\over 2}^-$ or ${1\over
  2}^-$ states (e.g. for the ground state and excited band head at
about 2.8~MeV, respectively, in $^9$Be), second-order derivatives in
the potential $V$ must enter. In the latter case, one also needs to
employ a potential that breaks spherical symmetry down to axial
symmetry, thus lifting the four-fold degeneracy of a $p_{3/2}$ orbital
in the body-fixed frame. The considerable number of low-energy
coefficients then requires that a significant amount of data is
available. In practice, one would like to adjust to scattering data
(and make predictions for spectra and transitions), but those are
rare.  Odd-mass neutron-rich isotopes of magnesium, for instance, are
expected to have ${5\over2}^+$ ground states. This would require us to
carry the expansion~(\ref{Veft}) to even higher order, and the
scarcity of data in rare isotopes would prohibit us to follow such an
EFT approach. It is, therefore, more practical to assume that the
Hamiltonian~(\ref{Hpsi}) is already in diagonal form, with low-energy
eigenstates as given in Eq.~(\ref{Hferm2}) fitted to the data. In
other words, we take the single-particle energies of the fermion from
data. This is somewhat similar in spirit to halo EFT~\cite{hammer2017}
where each state of the core is represented as a separate field and
simply adjusted to data.

The resulting EFT involves -- in leading order -- terms of order $\xi$
and $\Omega$. Subleading corrections are suppressed by factors of
$\xi/\Lambda$ (or $\xi/\Omega$ provided that $\xi\ll \Omega$
holds). This EFT does not provide us with an expansion in powers of
$\Omega / \Lambda$, because we do not construct such an expansion of
the potential $V$.


\subsection{Nucleon-rotor interactions} 
\label{sec:gauge}

We deal with emergent symmetry breaking. Thus, the nucleon can couple
to the rotor only derivatively, i.e., via the angular velocity
$\mathbf{v}$. All terms allowed by the symmetries must be considered.
At face value, the resulting velocity-dependent couplings are well
known. They involve -- in the body-fixed frame -- Coriolis forces. 
However, the essential physical
argument for the couplings is more subtle and profound. The coupling
terms are gauge couplings that involve Berry phases (or geometrical
phases). Such phases occur in many quantum
systems~\cite{simon1983,berry1984}. While originally conceived for
systems that undergo a time-dependent adiabatic motion, they may also
occur where ``fast'' degrees of freedom have been removed or
integrated out, and where one is only interested in the remaining
``slow" degrees of freedom~\cite{kuratsuji1985}. A well-known example
from molecular physics is the Born-Oppenheimer approximation. Here,
Berry phases and the corresponding gauge potentials enter the dynamics
of the nuclei of the molecule once the electronic degrees of freedom
have been removed. That leads to the molecular Aharonov-Bohm
effect~\cite{mead1979,wilczek1984}. For the diatomic molecule, some
details are presented in
Refs.~\cite{jackiw1986,bohm1992,rho1994}. In the present case, the fact
that the nucleon is much faster than the slowly rotating core shows
that gauge potentials play a role. Likewise, gauge potentials
are a general feature of systems where a separation between rotational
and intrinsic degrees of freedom is being made, and different
conventions for this separation differ by gauge
transformations~\cite{littlejohn1997}. 

The non-linear realization of rotational invariance requires that, in
the body-fixed system, we have to replace the time derivative
$i\partial_t$ in the Lagrangian~(\ref{Lpsi}) by the covariant
derivative~\cite{papenbrock2011} (see Appendices~\ref{sec:cova} and
\ref{sec:coset})
\ba
\label{cov-diff}
iD_t &\equiv& i\partial_t +\dot{\phi}\cos\theta \left[ J_{z'}, \cdot
  \right] \ .
\ea
Here, the commutator's second argument is left open. The last term of
the covariant derivative accounts for Coriolis effects in the
body-fixed system. It is present even if the Lagrangian in the
body-fixed system does not depend on time explicitly, i.e., even if
the partial time derivative vanishes. In the Lagrangian~(\ref{Lpsi})
that yields
\begin{align}
\label{cov-diff2}
&\int{\rm d}^3\mathbf{x}' \hat{\Psi}^\dagger(\mathbf{x}') iD_t
\hat{\Psi}(\mathbf{x}') \nonumber\\ &= \int{\rm d}^3\mathbf{x}'
\hat{\Psi}^\dagger(\mathbf{x}') i\partial_t \hat{\Psi}(\mathbf{x}') +
\mathbf{v}\cdot\left(\mathbf{e}_\phi \cot\theta \hat{K}_{z'} \right)
\ .
\end{align}
We have factored out the angular velocity $\mathbf{v}$, see
Eq.~(\ref{velocity}), and we have used Eq.~(\ref{nJ1}). This naturally
introduces the gauge potential
\ba
\label{Aabel}
\mathbf{A}_{\rm a}(\theta, \phi) \equiv  \mathbf{e}_\phi \cot\theta
\hat{K}_{z'}
\ea
which couples the rotor to the nucleon via
$\mathbf{v}\cdot\mathbf{A}_{\rm a}$. Here we borrow the expression
gauge potential from electrodynamics. In the parlance of differential
geometry, the field $\mathbf{A}_{\rm a}$ is a connection. The term
$\mathbf{v}\cdot\mathbf{A}_{\rm a}$ scales as $\xi$, i.e. the gauge
potential is dimensionless and of order one. Thus, the gauge term is
as important as the Lagrangian~(\ref{Lrot}) of the free rotor and
enters in leading order.

The gauge potential~(\ref{Aabel}) is singular at the north and south
poles of the unit sphere. Single-valuedness of the wave function for
the rotor requires that the eigenvalues $K$ of $\hat{K}_{z'}$ be
integer or half integer. That is obviously guaranteed for the fermion
for which $K = \pm {1\over 2}$, $\pm {3\over 2}$, $\cdots$. We compute
the corresponding magnetic field (or the Berry curvature in
differential geometry) and find
\ba
\label{Babel}
\mathbf{B}_{\rm a}(\theta, \phi)  &\equiv& \nabla_\Omega \times
\mathbf{A_{\rm a}} =  -\mathbf{e}_{r} \hat{K}_{z'} \ .
\ea
This is the field of a Dirac monopole on the unit sphere and clearly
exhibits spherical symmetry~\cite{fierz1944,wu1976}, in contrast to
the gauge potential~(\ref{Aabel}) whose rotational invariance is not
obvious. As shown in App.~\ref{sec:gaugetrafo}, the effect of a
rotation on the gauge potential~(\ref{Aabel}) can be reversed by a
gauge transformation.

To see that the field~(\ref{Babel}) is indeed a monopole field we take
a detour and consider a sphere of radius $R$ of the size of the
nucleus, with angular and radial coordinates $(\theta,\phi)$ and
$\rho$, respectively. We neglect excitations of the sphere in the
radial direction as these relate to vibration with energies at the
breakdown scale and put $\rho = R$. Then angular velocities and the
vector potential~(\ref{Aabel}) are multiplied with $R$ and $R^{-1}$,
respectively. The usual differential operator $\mathbf{e}_r\partial_R
+ R^{-1}\nabla_\Omega$ then shows that we deal indeed with a monopole
field.

The gauge potential~(\ref{Aabel}) is intimately linked to the geometry
of the sphere, i.e. the coset space SO(3)/SO(2).  To see this, we
consider a sequence of three rotations (around space-fixed axes) that
take the rotor from a point $A$ on the unit sphere to a point $B$,
then from $B$ to a point $C$, and finally from point $C$ back to the
point $A$.  We assume that the three rotations are around three
distinct axes. This ensures that the triangle $ABC$ on the sphere has
a finite solid angle (or area). It is clear that at least two of the
three rotations will also induce rotations of the fermion field around
the body-fixed $z'$-axis. While the rotor has returned to its original
configuration after the sequence of the three rotations, the fermion's
configuration has been changed by a finite rotation around the
body-fixed $z'$-axis.  Inspection shows that the rotation angle,
i.e. the phase acquired by the fermion with spin projection $K$, is
equal to the solid angle of the enclosed loop times the spin
projection $K$. Dynamically this phase is acquired because of the 
monopole magnetic field.

Another gauge coupling, permitted in the framework of our EFT, is
\ba
\label{gaugecoup2}
g \mathbf{v}\cdot\left(\mathbf{e}_{r}\times \mathbf{K} \right) \ .
\ea
Here, $g$ is a dimensionless coupling constant. Its natural size is of
order unity, and the contribution~(\ref{gaugecoup2}) scales as $\xi$
and enters at leading order.  The corresponding gauge potential is
\ba
\label{Anon}
\mathbf{A}_{\rm n}(\theta,\phi)= g\mathbf{e}_{r}\times \mathbf{K}
= g\left(\mathbf{e}_\phi \hat{K}_{x'}-\mathbf{e}_\theta \hat{K}_{y'}
\right) \ .
\ea
This vector potential contains non-commuting operators and, therefore,
constitutes a non-Abelian gauge potential. The corresponding magnetic
field is
\ba
\mathbf{B}_{\rm n}(\theta, \phi)  &\equiv& \nabla_\Omega \times
\mathbf{A}_{\rm n} -i \mathbf{A}_{\rm n}\times\mathbf{A}_{\rm n}
\nonumber\\
&=& g\mathbf{e}_{r} \cot\theta \hat{K}_{x'}  + g^2\mathbf{e}_{r}
\hat{K}_{z'} \ .
\ea
Taken by itself, this magnetic field is not invariant under
rotations. However, the total gauge potential is 
\be
\label{Atot}
\mathbf{A}_{\rm tot}
= \mathbf{A}_{\rm a}+\mathbf{A}_{\rm n} \ ,
\ee
and the corresponding magnetic field
\ba 
\mathbf{B}_{\rm tot} &\equiv& \nabla_\Omega \times\mathbf{A}_{\rm
  tot} -i \mathbf{A}_{\rm tot} \times\mathbf{A}_{\rm tot}
\nonumber\\ &=& (g^2-1)\mathbf{e}_{r} \hat{K}_{z'}
\label{Btot}
\ea
is spherically symmetric. Again, we deal with a magnetic monopole.
However, in contrast to the field~(\ref{Babel}) its overall strength
is not quantized because the non-Abelian vector potential~(\ref{Anon})
exhibits no singularities on the unit sphere and the coupling $g$ can
therefore assume any real value. 

To show how this gauge term relates to the Berry phase we observe that
the states $|{\pm K}, \alpha \rangle$ have the same energy, see
Eq.~(\ref{Hferm2}). When the rotor moves along a closed loop on the
unit sphere, a general interaction would mix the two degenerate states
while transversing the loop. Thus, the initial and final fermion
states could differ by a unitary transformation. The non-Abelian gauge
potential~(\ref{Anon}) generates such a mixing for non-zero values of
the coupling $g$.

The choice of the gauge potentials is not unique. At any orientation
$(\theta,\phi)$ of the rotor's symmetry axis, the body-fixed
coordinate system is defined up to an arbitrary rotation around the
$z'$ axis. Thus, the intrinsic degrees of freedom (of the fermion in
our case) depend on a convention which is arbitrary. Different choices
of the body-fixed system lead to different expressions for the
covariant derivative and to different gauge
potentials~\cite{littlejohn1997}. These are related to each other by
gauge transformations. Details are given in Appendix~\ref{sec:gauge1}.

\section{Lagrangian and Hamiltonian}
\label{lagr}

\subsection{Leading- order terms}
\label{subsec:L2}

Collecting the leading-order results from the previous Sections, we
find that the Lagrangian is given by
\ba
\label{L2}
L &=& {C_0\over 2}\mathbf{v}^2 + \mathbf{v}\cdot \left( \mathbf{A}_{\rm a}
+ \mathbf{A}_{\rm n}\right)  +L_\Psi\nonumber\\
&=&{C_0\over 2}\left(\dot{\theta}^2 +\dot{\phi}^2\sin^2\theta\right)
+ g \left(\dot{\phi}\sin\theta \hat{K}_{x'} -\dot{\theta}\hat{K}_{y'}
\right) \nonumber\\
&& + \dot{\phi}\cos\theta \hat{K}_{z'} + L_\Psi \ .
\ea
Here, $L_\Psi$ is defined in Eq.~(\ref{Lpsi}). The Legendre transform
of the Lagrangian~(\ref{L2}) yields the Hamiltonian. For $L_\psi$ that
was done in Section~\ref{sec:fermion}. For the remaining variables the
transformation is less tedious than might appear at first sight. The
Lagrangian~(\ref{L2}) is a quadratic form in the velocities
$(\dot{\theta},\dot{\phi})$ and can be written as
\be
L = {1\over 2} \dot{\mathbf{q}}^T \hat{M} \dot{\mathbf{q}} + \mathbf{A}
\cdot \dot{\mathbf{q}} \ ,
\ee
where $^T$ denotes the transpose. That Lagrangian has the Legendre
transform
\be
H = {1\over 2} \left(\mathbf{q}-\mathbf{A}\right)^T \hat{M}^{-1}
\left(\mathbf{q}-\mathbf{A}\right) \ .
\ee
Here, $\hat{M}$ is a ``mass" matrix and $\hat{M}^{-1}$ denotes its
inverse. In the present case the canonical momenta are
\ba
\label{can_mom}
p_\phi  &\equiv& {\partial L\over\partial \dot{\phi}}  = C_0 v_\phi
\sin\theta + \cos\theta \hat{K}_{z'} +g\sin\theta \hat{K}_{x'} \ ,
\nonumber \\
p_\theta &\equiv& {\partial L\over\partial \dot{\theta}} = C_0
v_\theta -g \hat{K}_{y'} \ .
\ea
Here, we employed Eq.~(\ref{v_theta_phi}). The Hamiltonian becomes
\ba
\label{Ham}
H &=& H_\psi + \frac{1}{2C_0}\left(p_\theta +g\hat{K}_{y'}\right)^2
\nonumber\\
&&+\frac{1}{2C_0} \left({p_\phi\over\sin\theta}-\cot\theta\hat{K}_{z'}
- g\hat{K}_{x'}\right)^2 \ .
\ea
Here, the fermion Hamiltonian $H_\psi$ is given in Eq.~(\ref{Hpsi}).
The momentum $p_\phi$ is conserved because $\phi$ does not appear in
the Hamiltonian~(\ref{Ham}). Combining two of the canonical
momenta~(\ref{can_mom}) into
\be
\mathbf{p} = p_\theta \mathbf{e}_\theta +{p_\phi\over \sin\theta}
\mathbf{e}_\phi \ ,
\ee
we find
\be
\label{pv}
\mathbf{p} = C_0\mathbf{v} + \mathbf{A}_{\rm  a} +\mathbf{A}_{\rm n} \ .
\ee
Using that, we write the Hamiltonian~(\ref{Ham}) in compact form as
\ba
\label{ham2}
H = H_\psi+ {1\over 2 C_0}\left(\mathbf{p} - \mathbf{A}_{\rm  a} -
\mathbf{A}_{\rm n} \right)^2 = H_\psi + \frac{1}{2} C_0\mathbf{v}^2 \ .
\nonumber \\
\ea 
%

\subsection{Angular momentum}

Replacing the canonical momenta by the total angular momentum
simplifies the Hamiltonian and establishes the connection to
Eq.~(\ref{Hpr1}). In the present Subsection we introduce the total
angular momentum on an intuitive basis. A derivation based on
Noether's theorem is given in App.~\ref{sec:emmy}.

The total angular momentum
\be
\label{Itot}
\mathbf{I} = \mathbf{I}_{\perp} + \mathbf{I}_{z'}
\ee
is the sum of the angular momentum of the fermion,
\ba
\label{Ipsi}
\mathbf{I}_{z'} =  \mathbf{e}^\prime_{z}  \hat{K}_{z'} \ , 
\ea
which points in the direction of the symmetry axis, and that of the
rotor,
\ba
\label{Irot}
\mathbf{I}_{\perp}
&=& I_{x'}\mathbf{e}^\prime_{x} +  I_{y'}\mathbf{e}^\prime_{y} \nonumber \\
&=& p_\theta\mathbf{e}^\prime_{y} - \left( {p_\phi\over\sin\theta}  -
\cot\theta\hat{K}_{z'}\right)\mathbf{e}^\prime_{x} \nonumber \\
&=& \mathbf{r}\times\left(\mathbf{p}-\mathbf{A}_{\rm a}\right) \ .
\ea
which is perpendicular to it. In the last line of Eq.~(\ref{Irot}) we
have used Eq.~(\ref{pv}), see also Refs.~\cite{fierz1944,wu1976}. The
term $\mathbf{r} \times \mathbf{p}$ is the angular momentum
$C_0\mathbf{r}\times\mathbf{v}$ of the rotor. The gauge potential
$\mathbf{A}_a$ is not manifestly invariant under rotations but can be
made so via a gauge transformation~\cite{fierz1944}. That causes the
correction~(\ref{Ipsi}) in the direction of $\mathbf{e}^\prime_{z}$.
The equality
\be
\label{Iz}
I_{z}\equiv \mathbf{e}_z \cdot \mathbf{I} = p_\phi \ .
\ee
shows that the conserved momentum $p_\phi$ is the usual angular
momentum with respect to the space-fixed $z$ axis. We use
Eqs.~(\ref{Irot}) and (\ref{Itot}) to express the angular velocity in
terms of the angular momentum. That yields
\ba
\mathbf{v} = -{1\over C_0}\left(\mathbf{e}^\prime_{z}\times\mathbf{I}
+ \mathbf{A}_{\rm n}\right) \ .
\ea
Using that in the rotational energy $(C_0/2)\mathbf{v}^2$, we arrive
at the Hamiltonian
\ba
H &=&  H_\Psi + {g^2\over 2C_0}\left(\hat{K}_{x'}^2+\hat{K}_{y'}^2\right)
\nonumber\\
&&+\frac{\mathbf{I}^2-\hat{K}_{z'}^2}{2C_0}  +{g\over C_0}\left(I_{x'}
\hat{K}_{x'} +I_{y'}\hat{K}_{y'}\right) \ .
\label{Hfinal}
\ea
The term proportional to $g^2$ in Eq.~(\ref{Hfinal}) might be absorbed
into $H_\Psi$ (and then be dropped).  The rotational part displayed in
the second line of Eq.~(\ref{Hfinal}) corresponds to the rotor
model~(\ref{Hpr1}) for the special case of axial symmetry, i.e. for
$C_{x'} = C_{y'}$.

The square of the total angular momentum is given by
\ba
\label{Itot2}
\mathbf{I}^2 &=& p_\theta^2 +{1\over \sin^2\theta} \left(p_\phi -\cos
\theta \hat{K}_{z'}\right)^2 +\hat{K}_{z'}^2 \nonumber\\
&=& p_\theta^2 +{1\over \sin^2\theta} \left(p_\phi^2 -2p_\phi \cos
\theta \hat{K}_{z'} +\hat{K}_{z'}^2 \right) \ . 
\ea
Upon quantization this operator, its projection $I_{z'} =\hat{K}_{z'}$
onto the $z'$-axis [see Eq.~(\ref{Ipsi})], and its projection $I_z =
p_\phi$ onto the $z$-axis [see Eq.~(\ref{Iz})] form a commuting set of
operators. Details are presented in Appendix~\ref{sec:emmy}.

\subsection{Spectrum}

Simplifying the notation used in Eq.~(\ref{Hferm2}) we denote the
ground state of the fermionic part of the Hamiltonian~(\ref{Hfinal})
as $|K\rangle$. We calculate the eigenfunctions of the rotor part of
the Hamiltionian~(\ref{Hfinal}) by determining the eigenfunctions of
$\mathbf{I}^2$, $I_z$, and $I_{z'}$. For states $|{\pm K}\rangle$ we
have
\ba
I_{z'} |{\pm K}\rangle = \hat{K}_{z'}|{\pm K}\rangle={\pm K} |{\pm K}
\rangle \ .
\ea
The quantization proceeds as in Section~\ref{sec:dof}. The
eigenfunctions of $I_z = p_\phi = -i\partial_\phi$ are
\ba
I_z e^{-iM\phi}  = -M e^{-iM\phi} \ . 
\ea
The negative eigenvalue is chosen here to be consistent with chapter
4.2 of Ref.~\cite{varshalovich1988}. The eigenfunctions of the square
of the total angular momentum operator can be written either in terms
of Wigner $d$ functions or in terms of Wigner $D$ functions (see
chapter 4 of Ref.~\cite{varshalovich1988}). These are related by
\be
\label{wignerD}
D_{M,{M'}}^I(\phi,\theta,0) = e^{-iM\phi} d^I_{M, M'}(\theta)  \ .
\ee
For $I\ge |M|, |K|$ we have
\begin{align}
&\mathbf{I}^2 D_{M,{\mp K}}^I(\phi,\theta,0)  |{\pm K}\rangle \\
&= I(I+1) D_{M,{\mp K}}^I(\phi,\theta,0) |{\pm K}\rangle\ .
\end{align}
For the Hamiltonian~(\ref{Hfinal}) that implies
\begin{align}
\label{Hrot2}
&\frac{\mathbf{I}^2-\hat{K}_{z'}^2}{2C_0} D_{M,{\mp K}}^I(\phi,\theta,0)
|{\pm K}\rangle \nonumber \\
&= \frac{I(I+1)-K^2}{2C_0} D_{M,{\mp K}}^I(\phi,\theta,0)  |{\pm K}
\rangle \ .
\end{align}

Discrete symmetries of the rotor-plus-fermion system may select a
definite linear combination of the states $|{\pm K} \rangle$. These
share the absolute value $|K|$ and have the same energy $E_{|K|}$ [see
  Eq.~(\ref{Hferm2})]. Combining $E_{|K|}$ with Eq.~(\ref{Hrot2})
yields
\be
\label{ErgK>1/2}
E(I) =  E_{|K|} + \frac{I(I+1)-K^2}{ 2C_0} \ .
\ee

The term linear in $g$ of the Hamiltonian~(\ref{Hfinal}) couples
states $D^I_{M,{-K}} |K \rangle$ and $D^I_{M,{-K \mp 1}} | K\pm
1\rangle$. For most heavy nuclei where $\xi \ll \Omega$, the coupling
of states with different values of $|K|$ is of subleading order and
can be computed perturbatively. For $K = \pm 1 / 2$, however, the
interaction couples the degenerate states $D^I_{M, {\mp {1\over 2}}} |
{\pm {1\over 2}} \rangle$ and is, thus, of leading order $\xi$. For
this case, eigenfunctions and eigenvalues are worked out in
Appendix~\ref{sec:suppl}. The result for the eigenvalues,
\ba
\label{Eresult}
E(I,K) &=&  E_{|K|}+ \frac{I(I+1)-K^2}{ 2C_0}\nonumber\\
&-&{g\over C_0} \delta_{|K|, {1\over 2}}(-1)^{I+{1\over 2}}\left(I +
{1\over 2} \right) \ ,
\ea
agrees with Eq.~(\ref{spectrum}) when we express the constants
$E_{|K|}$, $C_0$, and $g$ in terms of $E_0$, $A$, and $a$. The last
term in Eq.~(\ref{Eresult}) is known as the signature
splitting. Equation~(\ref{Eresult}) shows that for $g \neq 0$ the
spectrum changes by about $g\xi$. For $|g|\gg 1$, the spectrum would
resemble a rotational band only for $I\gtrsim |g|$. This confirms that
the natural size of $g$ is of order unity.

States that differ by one unit in $K$ can also be coupled strongly by
the term linear in $g$ in the Hamiltonian~(\ref{Hfinal}) provided
$\Omega \approx \xi$. In that case, the spectrum can be calculated
analytically using as a basis the eigenstates obtained for $g = 0$ and
taking account only of the two bandheads.  The diagonalization of the
$4\times4$ matrix spanned by the states $|{\pm K}\rangle$ and $|{\pm
  (K+1)}\rangle$ yields the eigenvalues~\cite{kerman1956}
\ba
\label{Ecomplete}
E(I,K,K+1) &=& {1\over 2}\left[E(I,K) +E(I,K+1)\right]
\nonumber\\ &\pm&{1\over 2} \bigg\{\left[E(I,K) - E(I,K+1)\right]^2
\nonumber\\ &&+4{\tilde{g}^2\over
  C_0^2}\left[I(I+1)-K(K+1)\right]\bigg\}^{1\over 2}
\ .\nonumber\\
\ea
The energies $E(I,K)$ are given by Eq.~(\ref{Eresult}), and
$\tilde{g}\equiv g\langle K|\hat{K}_{-1} |K{+1}\rangle$ is a
low-energy constant. The sign on the right-hand side of
Eq.~(\ref{Ecomplete}) has to be chosen such that the energies $E(I,K)$
and $E(I,K+1)$ for the bands with quantum numbers $K$ and $K+1$,
respectively, are obtained as $g\to 0$. In nuclei such as
$^{105,107}$Mo, groups of more than two band heads are closely spaced
and strongly coupled. In such cases, a Hamiltonian matrix of larger
dimension needs to be diagonalized.

We discuss our results. For $g = 0$, the total angular momentum
$\mathbf{I}^2$ and its projections $I_z$ and $I_{z'}$ onto the space-
and body-fixed $z$-axes, respectively, commute with each other and
with the Hamiltonian. The spectrum is given by
Eq.~(\ref{ErgK>1/2}). The nucleus is axially symmetric because
$I_{z'}$ is conserved. For finite $g$, the projection of the angular
momentum onto the rotor's symmetry axis is not conserved because the
Abelian and non-Abelian gauge potentials do not commute. According to
the rules for power counting, the term linear in $g$ (the ``Coriolis
term'') is of leading order. Nevertheless, the impact of the Coriolis
term on the spectrum depends very much on the nucleus under
consideration. In a band with band-head spin $K$ this term contributes
of the order $\xi(\xi/\Omega)^{K-1/2}$.  Thus, it is only of leading
order for a rotational band with $K = 1/2$. However, the Coriolis term
also couples bands that differ in $|K|$ by one
unit. Equation~(\ref{Ecomplete}) shows that the Coriolis term is of
leading order only if $\tilde{g}\equiv g\langle K|\hat{K}_{-1}
|K{+1}\rangle$ is sufficiently large, i.e. of order unity. In
practice, this is mostly expected if two band heads that differ in
spin by one unit are closely spaced in energy. Here ``close" means
that the spacing is not of the typical fermion scale $\Omega$ but
rather of the rotational scale ~$\xi$. In the presence of the Coriolis
term, $I_{z'}$ is not a conserved quantity anymore, and the odd-mass
nucleus exhibits triaxial deformation. We illustrate this behavior
below for $^{187}$Os. From the point of view of our EFT, triaxiality
in odd-mass nuclei thus depends on the spins of band heads and on
their separation in energy.

\subsection{Next-to-leading order corrections}

The leading-order Hamiltonian~(\ref{Hfinal}) contains contributions
that scale as $\xi$ and/or $\Omega$.  Out of the many terms quadratic
in both $\mathbf{v}$ and $\mathbf{K}$ that one can write down using
$\mathbf{v}$, $\mathbf{K}$, and $\mathbf{e}^\prime_{z}$, the
following combinations are linearly independent and compatible with
the symmetries:
\ba
\label{L-NLO}
L_{\rm 1a} &=& {g_{a}\over 2} \mathbf{v}^2 \left(K_{x'}^2 +K_{y'}^2\right)
\ , \nonumber\\
L_{\rm 1b} &=& {g_{b}\over 2} \mathbf{v}^2 K_{z'}^2  \ , \nonumber\\
L_{\rm 1c} &=& {g_{c}\over 2} \left(\mathbf{v}\cdot \mathbf{K}\right)^2 \ .
\ea
The natural assumption is that $g_{a,b,c}\sim\Lambda^{-1}$. Then, the
contributions of $L_{\rm 1a,b,c}$ scale as $\xi^2/\Lambda$, which is a
factor $\xi/\Lambda$ smaller than the leading-order
Lagrangian~(\ref{L2}). The next-to-leading order terms~(\ref{L-NLO})
are still quadratic in the velocities. After adding these terms to the
Lagrangian~(\ref{L2}) we can, therefore, perform the Legendre
transform as outlined in Subsection~\ref{subsec:L2}, but invert the
mass matrix perturbatively. The calculation is done in
Appendix~\ref{sec:suppl}. The resulting Hamiltonian is
\be
H = H_{\rm LO}+ H_{\rm NLO} \ , 
\ee
with the leading-order Hamiltonian $H_{\rm LO}$ as in
Eq.~(\ref{Hfinal}) and with
\ba
\label{Hnlo}
H_{\rm NLO} &=&  {1\over 2C_0}\left(\mathbf{N}^T \hat{C} \mathbf{N}
+ \mathbf{N}^T \hat{G} \mathbf{N} \right) \ .
\ea
The dimensionless operators $\hat{C}$ and $\hat{G}$ are all of order
$\xi / \Lambda$ and depend on bilinear combinations of the fermion
operators $\hat{K}_{x'}$, $\hat{K}_{y'}$, and $\hat{K}_{z'}$. In
Eq.~(\ref{Hnlo}) we also used
\ba
\mathbf{N}&\equiv& 
\left(\begin{array}{c}
I_{y'}\\
I_{x'} \end{array}\right)
+g\left(\begin{array}{c}
\hat{K}_{y'}\\
\hat{K}_{x'} \end{array}\right)
 \ .
\ea
The matrix $\hat{C}$, due to $L_{\rm 1a,b}$, is diagonal in the
eigenstates of the leading-order Hamiltonian~(\ref{Hfinal}). Thus, the
first term on the right-hand side of Eq.~(\ref{Hnlo}) adds a
fermion-state dependent correction of order $\xi^2/\Lambda$ to the
moment of inertia. It causes the moments of inertia of rotational
bands in odd mass nuclei to deviate somewhat from the moment of
inertia for the ground-state band of the even-even rotor. The
correction can be compared to the smaller variations of order
$(\xi^3/\Lambda^2)$ that occur in even-even
nuclei~\cite{zhang2013}. The matrix $\hat{G}$ (due to $L_{\rm 1c}$) in
the second term is traceless and mixes fermion states that differ in
quantum numbers $K_{z'}$ by two units. In particular, this term
modifies the rotational spectra of $|K_{z'}| = 3 / 2$ band heads.

How does our approach compare with a treatment that would use
full-fledged quantum field theory? While in the derivation of the EFT
we employed velocities and canonical momenta, the solution of the
Hamiltonian became simple because we introduced angular momenta. In
the gauge we used the eigenfunctions are the Wigner $D$
functions~(\ref{wignerD}). These can be written as infinite sums of
spherical harmonics, i.e. of the ``free'' solutions of the even-even
rotor. We are convinced that using $\nabla_\Omega \to
\nabla_\Omega-i\mathbf{A}_{\rm tot}$, gauging the quantum-field theory
Lagrangian~(\ref{QFT-L} with the gauge potential~(\ref{Atot}), and
using field-theoretical tools such as Feynman diagrams, would yield
the same result. Then, the ``free'' rotor would scatter via infinite
loops with the fermion, with vertices due to the gauge coupling.

\section{Applications}
\label{sec:apps}

In the previous Section we have shown that in leading order, the EFT
for odd-mass deformed nuclei recovers the results of the (axially
symmetric) particle-rotor model. While that model is well known, with
numerous applications to be found in
textbooks~\cite{rowe1970,eisenberg,bohr1975,iachello,rowe2010} and in
the literature, the EFT provides us, in addition, with a systematic
approach to subleading corrections and to estimates of the uncertainty
of EFT predictions~\cite{furnstahl2014c}. We illustrate that point,
following arguments made previously for even-even deformed nuclei with
axial symmetry~\cite{coelloperez2015} and for vibrational exitations
in heavy nuclei~\cite{coelloperez2015b,coelloperez2016}.

\subsection{$^{239}$Pu}

\begin{figure}[bt]
\includegraphics[width=0.45\textwidth]{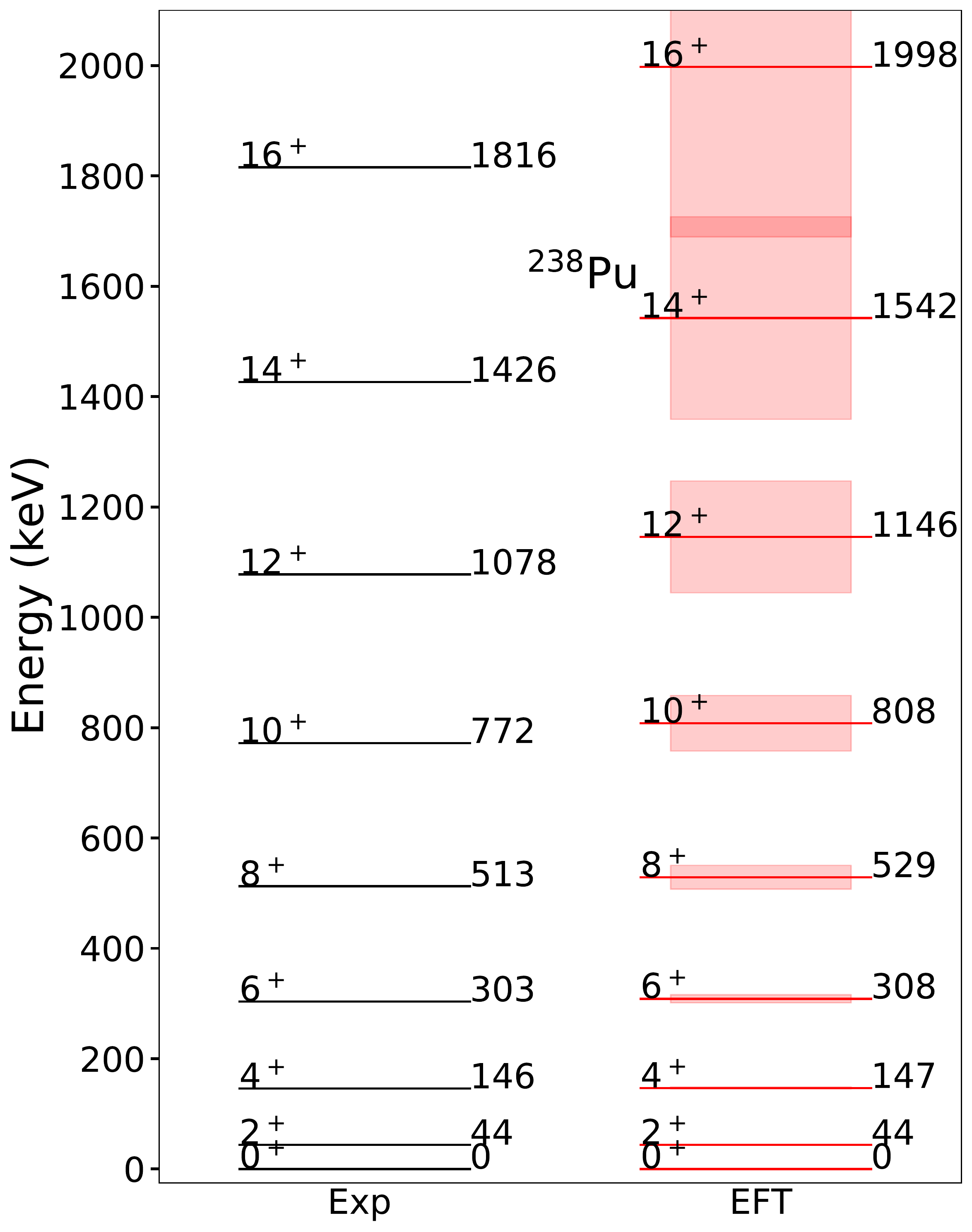}
\caption{(Color online) Levels of the ground-state rotational band in
  $^{238}$Pu, with spin/parity and energy as indicated, from data
  (left, black) are compared to EFT predictions (red, right) at
  leading order [${\cal O}(\xi)$] with uncertainty estimates (shaded
  red areas). }
\label{fig:pu238EFT}
\end{figure}

Within the EFT the nucleus $^{239}$Pu is described as a neutron
attached to $^{238}$Pu. Inspection of the low-lying states of
$^{238}$Pu in Fig.~\ref{fig:pu238} shows that the low-energy scale is
$\xi\approx 44$~keV and the breakdown scale is $\Lambda\approx
600$~keV.  This is probably too conservative an estimate for the
breakdown scale of the ground-state band in $^{238}$Pu, because the
lowest band head with positive parity occurs at 941~keV. Thus, for a
description of the ground-state band, $\xi/\Lambda\approx 1/21$ is
probably a more accurate estimate for the the expansion
parameter. Adjusting the low-energy constant $C_0$ to the energy of
the $2^+$ state yields $1/(2C_0)=7.35$~keV. The leading-order EFT
predictions~\cite{papenbrock2011,coelloperez2015} for the ground-state
rotational band are levels at energies
\be
E_{\rm LO}(I) = \frac{I(I+1)}{2C_0} \left[1+ {\cal O}\left({\xi^2\over
    \Lambda^2}\right)I(I+1)\right] \ .
\ee
Here, we included the EFT uncertainty
estimate~\cite{papenbrock2011,coelloperez2015}. Figure~\ref{fig:pu238EFT}
compares the EFT results to data. For the uncertainty estimate we used
${\cal O}\left({\xi^2\over\Lambda^2}\right)=0.25 (\xi/\Lambda)^2$,
where the factor 0.25 is determined empirically.

\begin{figure}[bt]
\includegraphics[width=0.45\textwidth]{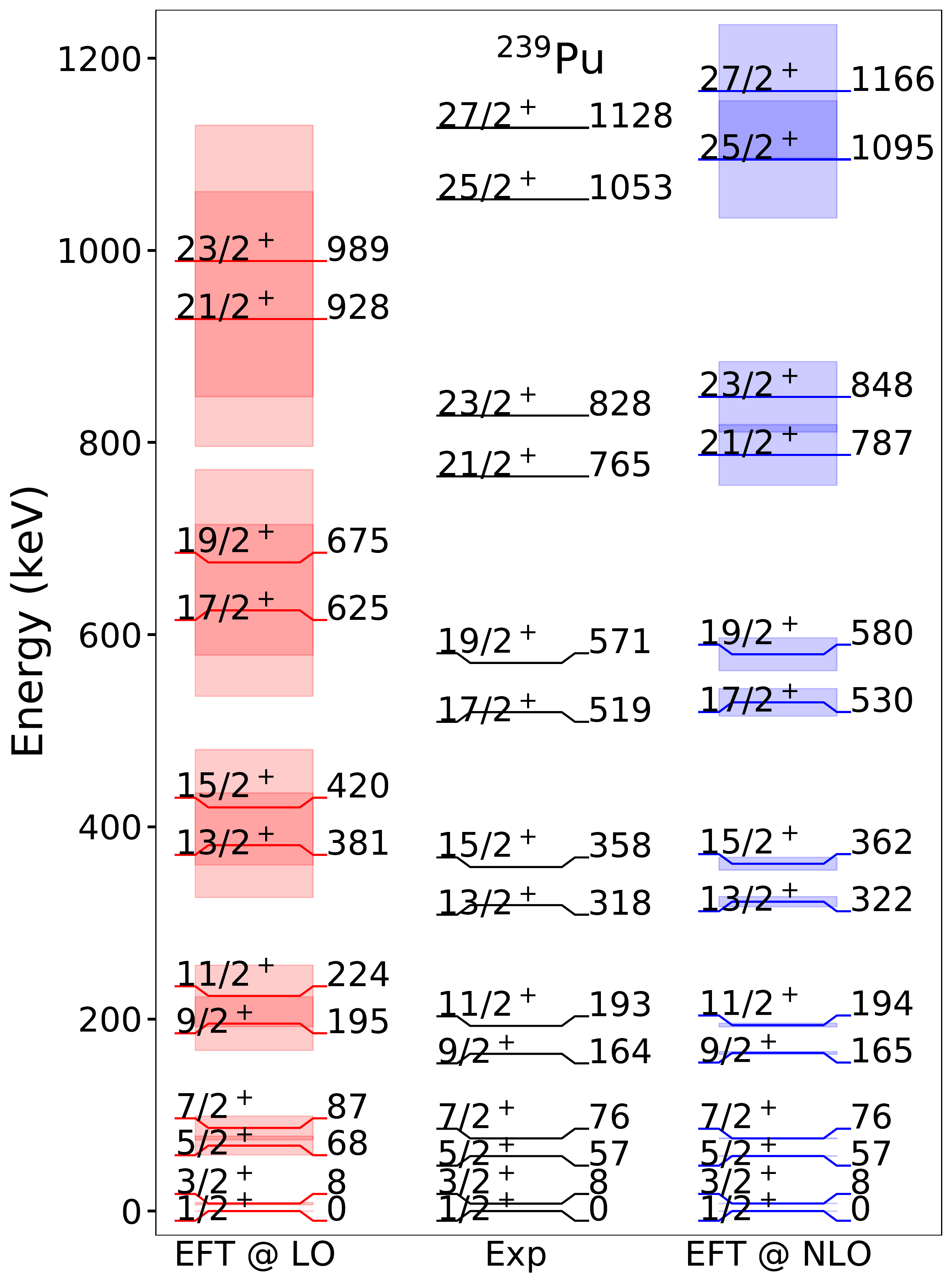}
\caption{(Color online) Levels of the ground-state rotational band in
  $^{239}$Pu, with spin/parity and energy as indicated, from data
  (center, black) are compared to EFT predictions at leading order
  (red, left) and at next-to-leading order (blue, right) with
  uncertainty estimates (shaded areas). }
\label{fig:pu239EFT}
\end{figure}

In leading order, the rotational constant of the nucleus $^{239}$Pu is
the same as for $^{238}$Pu. We only have to adjust the constant $g$ in
Eq.~(\ref{Eresult}) to describe the ground-state band. A fit to the
first excited state in this nucleus yields $g=-0.642$. The resulting
ground-state band is shown in the left part of Fig.~\ref{fig:pu239EFT}
and compared to data in the center. At leading order, the rotational
constant has a relative uncertainty of ${\cal O}(\xi/\Lambda)$, as
reflected by the blue shaded areas. For the displayed uncertainties,
we used the conservative estimate $\xi/\Lambda=1/14$ and ${\cal
  O}(\xi/\Lambda) = 2\xi/\Lambda$, with the factor of 2 determined
empirically.

A next-to-leading order fit to the energies $E(I,1/2)$ of
Eq.~(\ref{Eresult}) is shown in the right part of
Fig.~\ref{fig:pu239EFT}. Here, we adjusted both $C_0$ and $g$ in
Eq.~(\ref{Eresult}), finding $1/(2C_0)=6.257$~keV and $g=-0.579$. We
note that the change of $C_0$ by about a factor of $2\xi/\Lambda$ is
consistent with EFT expectations. At next-to-leading order, relative
energy uncertainties are estimated as $2C_0 E(I,1/2) {\cal
  O}(\xi^2/\Lambda^2)$ with ${\cal O}(\xi^2/\Lambda^2) = (0.25
\xi/\Lambda)^2$. As before, the factor 0.25 is determined empirically.

We see that the EFT yields an accurate (it agrees with the
data within the uncertainties) and increasingly precise (as more
orders are included) description of the ground-state rotational band
of $^{239}$Pu. Furthermore, the low-energy constants are not merely
fit parameters, but the size of subleading corrections can be
estimated from the empirical values of the low-energy scale $\xi$ and
the breakdown scale $\Lambda$. Similar results can also be obtained
for the other rotational bands displayed in Fig.~\ref{fig:pu239}.

\subsection{$^{187}$Os}

\begin{figure*}[tb]
\includegraphics[width=0.95\textwidth]{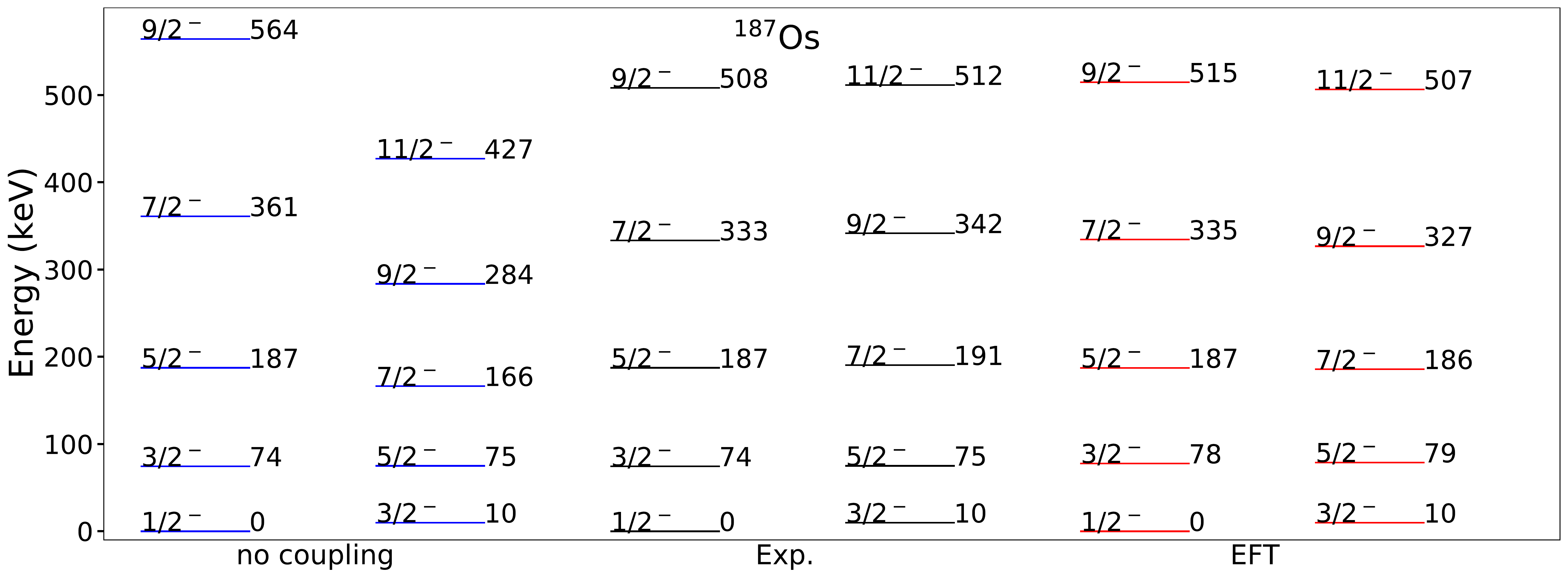}
\caption{(Color online) Levels of the two lowest-lying rotational
  bands in $^{187}$Os, with spin/parity and energy as indicated.
  Center (black): Data. Left (blue): Results obtained by fitting
  energies of both bands but neglecting the Coriolis coupling. Right
  (red): EFT fits with predictions at leading plus next-to-leading
  order. The relative EFT uncertainties (not shown) are about
  $2\xi^2/\Lambda^2\approx 7$\%.}
\label{fig:os187EFT}
\end{figure*} 

In most odd-mass nuclei, the Coriolis term~[last term in
  Eq.~(\ref{Hfinal})] that couples different rotational bands enters
only perturbatively, because band heads that differ in $K$ by one unit
are usually an energy $\Omega\gg\xi$ apart. However, in nuclei with
closely spaced band heads, the Coriolis term is of leading
order. Among these are the light nucleus $^9$Be, the nuclei $^{49}$Cr
and $^{49}$Mn, $^{105,107}$Mo, $^{187}$Ir, and $^{187}$Os. We
illustrate our results for the well-studied nucleus
$^{187}$Os~\cite{malmskog1971,morgen1973,sodan1975}. The $K^\pi=1/2^-$
ground state exhibits a rotational band with a low-energy constant
$C_0^{-1}\approx 47$~keV. The first excited $K^\pi=3/2^-$ band head is
only separated by $\Omega\approx 10$~keV. Thus, we have
$\xi\sim\Omega$, the two bands in question are coupled by the Coriolis
term, and Eqs.~(\ref{Ecomplete}) must be employed.

\begin{figure}[bt]
\includegraphics[width=0.48\textwidth]{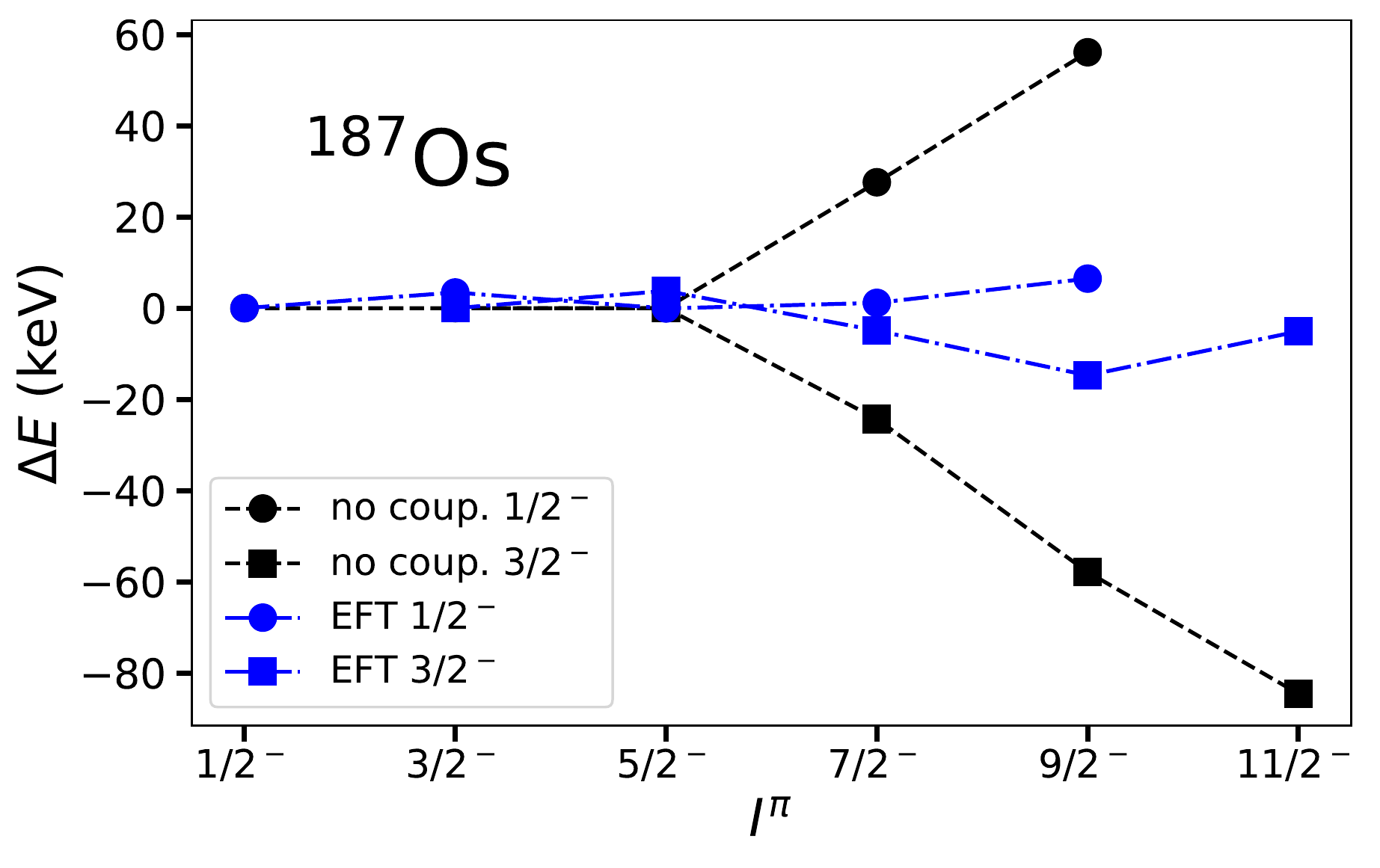}
\caption{(Color online) Energy differences between theory and data for
  the two lowest-lying rotational bands in $^{187}$Os as a function of
  spin/parity. Results obtained by fitting energies of both bands but
  neglecting the Coriolis coupling are shown in black, and EFT results
  in blue. Circles and squares mark the rotational states on top of
  the $I^\pi=1/2^-$ and $3/2^-$ band heads, respectively.}
\label{fig:os187diff}
\end{figure}

The relevant scales are as follows. The even-even nucleus $^{186}$Os
exhibits a ground-state rotational band with a $2^+$ state at 137~keV;
the excited $2^+$ band head at 770~keV sets the breakdown scale
$\Lambda$ of this rotor. The ratio of the energies of the two lowest
$2^+$ states is $\xi/\Lambda\approx 1/6$. In a first step we neglect
the coupling between the $K^\pi=1/2^-$ and $K^\pi=3/2^-$ bands in
$^{187}$Os. Adjusting a total of five parameters [$C_0$, $E_{1/2}$,
  and $g$ in Eq.~(\ref{Eresult})] to the lowest three states for
$K^\pi=1/2^-$ and fitting separately $C_0$ and $E_{3/2}$ to the two
states of the $K^\pi=3/2^-$ band yields the two rotational bands shown
in left part of Fig.~\ref{fig:os187EFT}. Here, the highest two (three)
states of the $K^\pi=1/2^-$ ($K^\pi=3/2^-$) band, respectively, are
predictions. The results are to be compared to the data shown in the
center. Also shown in the right part are the EFT predictions obtained
by adjusting the five parameters $C_0$, $E_{1/2}$, $E_{3/2}$ $g$, and
$\tilde{g}$ in Eq.~(\ref{Ecomplete}) simultaneously to the lowest
three states of both bands. Given the same number (five) of low-energy
constants, the improved accuracy obtained in the second fit shows the
need to include the Coriolis coupling. Comparing the results to the
data we infer that relative EFT uncertainties are about
$2\xi^2/\Lambda^2\approx 7$\%. Figure~\ref{fig:os187diff} shows the
energy differences beween theory and data for both bands using EFT
(blue) and neglecting the coupling between the bands (black). We see
that approach that neglects the coupling between the bands loses
accuracy as soon as one considers states that were not fitted.

\section{Summary}
\label{sec:sum}

We have developed an effective field theory for deformed odd-mass
nuclei. In this approach, the odd nucleon experiences an
axially-symmetric potential in the body-fixed frame of the even-even
deformed nucleus (a rotor). The power counting is based on the
separation of scales between low-lying rotational degrees of freedom
on the one hand and both, higher-lying nucleonic excitations and
intrinsic excitations of the even-even nucleus, on the other. In
leading order, the nucleon is coupled to the rotor via gauge
potentials. Actually, the non-Abelian gauge potential is a truly
first-order term only for $K = 1 / 2$ band heads or when band heads
with $K$ quantum numbers that differ by one unit of angular momentum,
are close in energy. In the latter case, the gauge potential induces
triaxiality. That was shown by applying the EFT to $^{187}$Os. We have
shown how subleading contributions can be constructed systematically,
and how these may be used to improve the spectrum and/or to estimate
theoretical uncertainties. The EFT developed in this paper presents a
model-independent approach to the particle-rotor system that is
capable of systematic improvement.

\begin{acknowledgments}
This work has been supported by the U.S.  Department of Energy under
grant No. DE-FG02-96ER40963  and under contract
DE-AC05-00OR22725 with UT-Battelle, LLC (Oak Ridge National
Laboratory).
\end{acknowledgments}

\newpage
\appendix

\section{Overview}

Appendix~\ref{trans} presents details regarding transformation
properties under rotations. In App.~\ref{sec:cova} we derive the
expression for the covariant derivative. Appendix~\ref{sec:coset}
presents a more formal derivation of these properties based on the
coset approach. In App.~\ref{sec:gauge1} we discuss gauge potentials
and gauge transformations. App.~\ref{sec:suppl} presents details
regarding the derivation of the spectrum and subleading corrections.

\section{Transformation properties under rotations}
\label{trans}

In this Appendix we use infinitesimal rotations and apply Noether's
theorem to derive the expressions in Section~\ref{lagr} for the total
angular momentum, both in the space-fixed and in the body-fixed system.

An infinitesimal rotation changes the angles $(\theta,\phi)$ to
$(\theta+\delta\theta,\phi+\delta\phi)$. That moves the symmetry axis
of the rotor into a new direction, and it induces a rotation of the
axes $\mathbf{e}^\prime_{x}$ and $\mathbf{e}^\prime_{y}$ around the
rotors new symmetry axis by an angle $\delta\omega$. In the following
two subsections, we relate the infinitesimal angles $(\delta\theta,
\delta\phi,\delta\omega)$ to the parameters of a general infinitesimal
rotation. We do so for rotations around the axes of the space-fixed
system and for rotations around the body-fixed axes. We use that at
the point $(\theta+\delta\theta,\phi+\delta\phi)$, the body-fixed
basis vectors are
\ba
\label{app2}
\mathbf{e}^\prime_{x} (\phi + \delta\phi, \theta + \delta \theta) &=&
\mathbf{e}^\prime_{x} (\phi, \theta) - \delta \theta
\mathbf{e}^\prime_{z}(\phi, \theta) \nonumber\\
&&+ \delta \phi \cos \theta \mathbf{e}^\prime_{y} (\phi, \theta) \ ,
\nonumber\\ 
\mathbf{e}^\prime_{y} (\phi + \delta\phi, \theta + \delta \theta) &=&
\mathbf{e}^\prime_{y} (\phi, \theta)  - \delta \phi \sin \theta
\mathbf{e}^\prime_{z}(\phi, \theta) \nonumber\\
&&- \delta \phi \cos \theta \mathbf{e}^\prime_{x} (\phi,\theta) \ ,
\nonumber\\
\mathbf{e}^\prime_{z} (\phi + \delta\phi, \theta + \delta \theta) &=&
\mathbf{e}^\prime_{z}
(\phi, \theta) + \delta \theta \mathbf{e}^\prime_{x} (\phi, \theta)
\nonumber\\
&&+\delta \phi \sin \theta \mathbf{e}^\prime_{y} (\phi, \theta) \ .
\ea

\subsection{Rotations around the space-fixed axes}
\label{rotsf}

A rotation by the vector $\delta \boldsymbol{\alpha} =
\delta\alpha_{x}\mathbf{e}_{x} + \delta\alpha_{y} \mathbf{e}_{y} +
\delta\alpha_{z}\mathbf{e}_{z}$ about infinitesimal angles $\delta
\alpha_k$, $k = x, y, z$ around the space-fixed axes changes the
body-fixed basis vectors $\mathbf{e}^\prime_{k}$, $k = x, y, z$ by
\ba
\label{app1}
\delta\boldsymbol{\alpha} \times \mathbf{e}^\prime_{x} &=&
\left(\delta\boldsymbol{\alpha} \cdot \mathbf{e}^\prime_{z}\right)
\mathbf{e}^\prime_{y}  
-\left(\delta\boldsymbol{\alpha} \cdot \mathbf{e}^\prime_{y}\right)
\mathbf{e}^\prime_{z}  \ , \nonumber\\
\delta\boldsymbol{\alpha} \times \mathbf{e}^\prime_{y} &=&  \left(
\delta\boldsymbol{\alpha} \cdot \mathbf{e}^\prime_{x}\right)
\mathbf{e}^\prime_{z} - \left(\delta\boldsymbol{\alpha} \cdot
\mathbf{e}^\prime_{z}\right) \mathbf{e}^\prime_{x}   \ , \\
\delta\boldsymbol{\alpha} \times \mathbf{e}^\prime_{z} &=& \left(
\delta\boldsymbol{\alpha} \cdot \mathbf{e}^\prime_{y}\right)
\mathbf{e}^\prime_{x} - \left(\delta\boldsymbol{\alpha} \cdot
\mathbf{e}^\prime_{x}\right) \mathbf{e}^\prime_{y}  \ . \nonumber
\ea
We equate the incremental changes of $\mathbf{e}^\prime_{z}(\phi +
\delta\phi, \theta + \delta \theta)$ on the right-hand side of the
last of Eqs.~(\ref{app2}) with the last line of Eq.~(\ref{app1}).
That yields
\ba
\label{trafo_theta_phi}
\left(
\begin{array}{c}
\delta\theta\\
\delta\phi
\end{array}\right) 
=\left[
\begin{array}{ccc}
-\sin \phi & \cos \phi & 0\\
-\cos \phi \cot \theta & - \sin \phi \cot \theta & 1
\end{array}\right] 
\left(
\begin{array}{c}
\delta \alpha_x\\
\delta \alpha_y\\
\delta \alpha_z
\end{array}\right) . \nonumber\\
\ea
The rotor's degrees of freedom clearly transform non-linearly, i.e.,
under the rotation by $\delta \boldsymbol{\alpha}$ they depend in a
nonlinear way on $(\phi, \theta)$. The rotated basis vector
$\mathbf{e}^\prime_{x}+\delta\boldsymbol{\alpha} \times
\mathbf{e}^\prime_{x}$ differs from the basis vector
$\mathbf{e}^\prime_{x}(\phi + \delta\phi, \theta + \delta \theta)$ by
a small rotation with the angle $\delta\omega$ around the rotor's
symmetry axis $\mathbf{e}^\prime_{z}(\phi + \delta\phi, \theta +
\delta \theta)$.  To determine $\delta\omega$ we compute the scalar
product
\begin{align}
\label{gamma}
\delta\omega &= \left[\mathbf{e}^\prime_{x}(\phi, \theta) +
  \delta\boldsymbol{\alpha} \times \mathbf{e}^\prime_{x}(\phi, \theta)
  \right]\cdot \mathbf{e}^\prime_{y} (\phi + \delta\phi, \theta +
\delta \theta)  \nonumber\\
&= {\cos \phi \over \sin \theta} \delta \alpha_x + {\sin \phi \over
  \sin \theta} \delta \alpha_y \ .
\end{align}

The rotation by the infinitesimal angle $\delta\omega$ around the
body-fixed $z'$-axis is induced by the operator $e^{-i\delta\omega
  J_{z'}}$. Under that transformation the spinor function
$\hat{\Psi}(\mathbf{x}')$, defined in Eq.~(\ref{spinor-s}) in the
body-fixed system, transforms as
\be
\label{psi-rot1}
\hat{\Psi}(\mathbf{x}') \to \hat{\Psi}(\mathbf{x}') + \delta
\hat{\Psi}(\mathbf{x}')
\ee
where 
\ba
\label{psi-rot}
\delta \hat{\Psi}(\mathbf{x}') &=& - i \delta\omega
\left[ J_{z'} , \hat{\Psi}(\mathbf{x}') \right] \ .
\ea
Collecting results from Eqs.~(\ref{trafo_theta_phi}), (\ref{gamma}),
and (\ref{psi-rot}) we find
\ba
\label{noether-mat1}
\left(
\begin{array}{c}
\delta\theta\\
\delta\phi\\
\delta\hat{\hat{\Psi}}(\mathbf{x}')
\end{array}\right) =\hat{T}_{\rm S} 
\left(
\begin{array}{c}
\delta \alpha_{x}\\
\delta \alpha_{y}\\
\delta \alpha_{z}
\end{array}\right) \ ,
\ea
where
\ba
\label{noether-mat}
\hat{T}_{\rm S} \equiv \left[
\begin{array}{ccc}
-\sin \phi & \cos \phi & 0\\
-\cos \phi \cot \theta & - \sin \phi \cot \theta & 1\\
-i {\cos\phi\over\sin\theta} \left[J_{z'} , \hat{\Psi}
  (\mathbf{x}')\right]
& -i{\sin\phi\over\sin\theta} \left[J_{z'} ,  \hat{\Psi}(\mathbf{x}')
  \right] & 0 \end{array}\right] \ . \nonumber\\
\ea
%

\subsection{Rotation around the body-fixed axes}
\label{rotbf}

A rotation by the vector $\delta \boldsymbol{\alpha}' =
\delta\alpha_{x'}\mathbf{e}^\prime_{x} + \delta\alpha_{y'}
\mathbf{e}^\prime_{y} + \delta\alpha_{z'}\mathbf{e}^\prime_{z}$ about
infinitesimal angles $\delta \alpha_{k'}$, $k' = x', y', z'$ around
the body-fixed axes changes the body-fixed basis vectors
$\mathbf{e}^\prime_{k'}$ by
\ba
\label{bf1}
\delta\boldsymbol{\alpha}' \times \mathbf{e}^\prime_{x} &=&
\delta\alpha_{z'} \mathbf{e}^\prime_{y} -\delta\alpha_{y'}
\mathbf{e}^\prime_{z} , \nonumber\\
\delta\boldsymbol{\alpha}' \times \mathbf{e}^\prime_{y} &=&
\delta\alpha_{x'} \mathbf{e}^\prime_{z} -\delta\alpha_{z'}
\mathbf{e}^\prime_{x} , \\
\delta\boldsymbol{\alpha}' \times \mathbf{e}^\prime_{z} &=&
\delta\alpha_{y'} \mathbf{e}^\prime_{x} -\delta\alpha_{x'}
\mathbf{e}^\prime_{y} . \nonumber
\ea
Equating the incremental change of $\mathbf{e}^\prime_{z}(\phi +
\delta\phi, \theta + \delta \theta)$ on the right-hand side of the
last of Eqs.~(\ref{app2}) with the last line of Eq.~(\ref{bf1}) gives
\ba
\label{bf-trafo_theta_phi}
\left(
\begin{array}{c}
\delta\theta\\
\delta\phi
\end{array}\right) 
=\left[
\begin{array}{ccc}
0 & 1 & 0\\
-{1\over\sin\theta} & 0 & 0
\end{array}\right] 
\left(
\begin{array}{c}
\delta \alpha_{x'}\\
\delta \alpha_{y'}\\
\delta \alpha_{z'}
\end{array}\right)  \ . 
\ea
The incremental rotation angle $\delta\omega'$ is given by the scalar
product of the rotated basis vector
$\mathbf{e}^\prime_{x}+\delta\boldsymbol{\alpha}' \times
\mathbf{e}^\prime_{x}$ and the basis vector $\mathbf{e}^\prime_{y}
(\theta+\delta\theta,\phi+\delta\phi)$,
\begin{align}
\label{bf-gamma}
\delta\omega'&=\left[\mathbf{e}^\prime_{x}(\theta,\phi) +
  \delta\boldsymbol{\alpha}' \times \mathbf{e}^\prime_{x}(\theta,\phi)
  \right] \cdot \mathbf{e}^\prime_{y} (\theta+\delta\theta,\phi +
\delta\phi) \nonumber\\
&= \delta\alpha_{x'} \cot\theta +\delta\alpha_{z'} \ .
\end{align}
That shows that a rotation by $\delta\boldsymbol{\alpha}'$ points the
body-fixed system into the new direction
$(\theta+\delta\theta,\phi+\delta\phi)$ and rotates the body fixed
system around its new axis $\mathbf{e}^\prime_{z}
(\theta+\delta\theta,\phi+\delta\phi)$ by the angle $\delta\omega'$.
The fermion wave function transforms as in Eqs.~(\ref{psi-rot1},
\ref{psi-rot}) but with $\delta \omega$ replaced by $\delta
\omega'$. Thus,
\ba
\label{bf-noether-mat1}
\left(
\begin{array}{c}
\delta\theta\\
\delta\phi\\
\delta\hat{\Psi}(\mathbf{x}')
\end{array}\right) 
=\hat{T}_{\rm B}
\left(
\begin{array}{c}
\delta \alpha_{x'}\\
\delta \alpha_{y'}\\
\delta \alpha_{z'}
\end{array}\right) \ , 
\ea
with
\ba
\label{bf-noether-mat}
\hat{T}_{\rm B}\equiv\left[
\begin{array}{ccc}
0 & 1 & 0\\
-{1\over\sin\theta} & 0 & 0\\
-i  \cot\theta \left[J_{z'} , \hat{\Psi}(\mathbf{x}') \right] & 0 &
-i \left[J_{z'} , \hat{\Psi}(\mathbf{x}') \right]
\end{array}\right] \ .\nonumber\\
\ea
%

\subsection{Noether's theorem and angular momentum}
\label{sec:emmy}

We use Noether's theorem~\cite{noether1918} and the results of
Subsections~\ref{rotsf} and \ref{rotbf} to obtain expressions for the
conserved quantities, i.e., the components of total angular momentum
in the space-fixed and in the body-fixed system, respectively. These
are used in Section~\ref{lagr} of the main text.

The theorem expresses invariants of the system in terms of partial
derivatives of the Lagrangian with respect to the velocities
$\dot{q}_\nu$ of the system. The Lagrangian is given by the second of
Eqs.~(\ref{L2}), with $L_\Psi$ defined in Eq.~(\ref{Lpsi}). The
velocities are $\dot{\theta} \equiv \dot{q}_1 $, $\dot{\phi} \equiv
\dot{q}_2$, and the time derivative $\partial_t
\hat{\Psi}(\mathbf{x}') \equiv \dot{q}(\mathbf{x}')$ of the fermion
wave function. The conserved quantities are the components of angular
momentum, expressed in terms of the transformation matrices of
Eqs.~(\ref{noether-mat}) and (\ref{bf-noether-mat}) and given by
\be
I_k = \sum_\nu \frac{\partial L}{\partial \dot{q}_\nu} \left[
  \hat{T}_{\rm S}\right]_{\nu k}
\label{noethersf}
\ee
for rotations around the space-fixed axes and 
\be
I_{k'} = \sum_\nu \frac{\partial L}{\partial \dot{q}_\nu} \left[
  \hat{T}_{\rm B}\right]_{\nu k'}
\label{noetherbf}
\ee
for rotations around the body-fixed axes. In the case of the velocity
$\dot{q}(\mathbf{x}')$, the summations on the right-hand sides of
Eqs.~(\ref{noethersf}) and (\ref{noetherbf}) actually involve an
integration over $\mathbf{x}'$. We use Eq.~(\ref{can_mom}), perform the
space integration over the matrix elements of $\hat{T}_{\rm S,B}$, and
use Eq.~(\ref{nJ1}). In the space-fixed system we find
\ba
I_{x} &=& -\sin \phi p_\theta - \cos \phi \cot \theta p_\phi +
\hat{K}_{z'}\frac{\cos \phi}{\sin \theta} \ , \nonumber \\
I_{y} &=&  \cos \phi p_\theta - \sin \phi \cot \theta p_\phi +
\hat{K}_{z'}\frac{\sin \phi}{\sin \theta} \ , \nonumber \\
I_{z} &=&  p_\phi \ .
\label{I-components}
\ea
In the body-fixed system we have
\ba
\label{I-components-bf}
I_{x'} &=& - \frac{p_\phi -\hat{K}_{z'}\cos\theta}{\sin\theta} \ ,
\nonumber\\
I_{y'} &=& p_\theta \ , \nonumber\\
I_{z'} &=&  \hat{K}_{z'}  \ .
\ea
The square of the total angular momentum, defined by the sum of the
squares of its components and calculated either in the space-fixed or
in the body-fixed system, in both cases is given by
Eq.~(\ref{Itot2}). Upon quantization, the
components~(\ref{I-components-bf}) do not fulfill the canonical
commutation relations as they are not generators of rotations. For a
discussion of unusual commutation relations we refer the reader to
Ref.~\cite{nauts2010}.

\section{Covariant derivative}
\label{sec:cova}

For the time derivative of a vector
$\mathbf{a}=a_{x'}\mathbf{e}^\prime_{x} +a_{y'}\mathbf{e}^\prime_{y}$
in the tangential plane of the two-sphere at $\mathbf{e}^\prime_{z}$,
we use
\ba
\dot{\mathbf{e}}_{x'} &=& -\dot{\theta}\mathbf{e}^\prime_{z}
+\dot{\phi}\cos\theta\mathbf{e}^\prime_{y} \ , \nonumber\\
\dot{\mathbf{e}}_{y'} &=& -\dot{\phi}\sin\theta \mathbf{e}^\prime_{z}
-\dot{\phi}\cos\theta \mathbf{e}^\prime_{x} \ ,
\ea
and have
\ba
\label{cov-diff1}
\dot{\mathbf{a}} &=& \left(\dot{a}_{x'} -a_{y'}\dot{\phi}
\cos\theta\right)\mathbf{e}^\prime_{x} +\left(\dot{a}_{y'} +a_{x'}
\dot{\phi}\cos\theta\right)\mathbf{e}^\prime_{y} \nonumber\\
&&-\left( a_{x'} \dot{\theta} + a_{y'} \dot{\phi}\sin\theta
\right)\mathbf{e}^\prime_{z} \ .
\ea
The projection of $\dot{\mathbf{a}}$ onto the tangential plane defines
the covariant derivative
\ba
\label{divDt}
D_t {\mathbf{a}}&\equiv& \left(\dot{a}_{x'} -a_{y'}\dot{\phi}
\cos\theta\right)\mathbf{e}^\prime_{x} +\left(\dot{a}_{y'} +a_{x'}
\dot{\phi}\cos\theta\right)\mathbf{e}^\prime_{y} \nonumber \\
&=& \partial_t \mathbf{a} -i \dot{\phi}\cos\theta J_{z'}\mathbf{a} \ .
\ea
The covariant derivative consists of the usual time derivative and a
rotation in the tangential plane, i.e., a rotation by
$\dot{\phi}\cos\theta$ around the $\mathbf{e}^\prime_{z}$ axis.

It is straightforward to generalize this argument to spin functions.
Let $\chi_{Sm}$ with spin $S$ and projection $m$ be the spin function
in the space-fixed system. A rotation to the body-fixed system yields
the helicity spin states
\be
\label{helicity}
\chi_{S \lambda}(\theta,\phi) = \sum_m D^S_{m \lambda}(\phi,\theta,0)
\chi_{Sm} ß .
\ee
These are quantized with respect to the body-fixed $z'$ axis, see
chapter 6.1.3 of Ref.~\cite{varshalovich1988}. The time derivative is
\ba
\dot{\chi}_{S \lambda}(\theta,\phi) &=& {\dot{\theta}\over 2}\sqrt{S(S+1)
  -\lambda(\lambda-1)} \chi_{S \lambda-1}(\theta,\phi) \nonumber\\
&-&{\dot{\theta}\over 2}\sqrt{S(S+1)-\lambda(\lambda+1)}
\chi_{S \lambda+1}(\theta,\phi) \nonumber\\
&-&i\dot{\phi}\sum_m mD^S_{m \lambda}(\phi,\theta,0) \chi_{Sm} \ .
\ea
Here we used formulas from chapter 4.9 of Ref.~\cite{varshalovich1988}.
We also find
\ba
\lefteqn{m D_{m\lambda}^S(\phi,\theta,0) = \lambda\cos\theta
  D_{m\lambda}^S(\phi,\theta,0)}\nonumber\\
&&-{\sin\theta\over 2}\sqrt{S(S+1)-\lambda(\lambda-1)}
D_{m\lambda-1}^S(\phi,\theta,0) \nonumber\\
&&-{\sin\theta\over 2}\sqrt{S(S+1)-\lambda(\lambda+1)}
D_{m\lambda+1}^S(\phi,\theta,0) \ .\nonumber\\
\ea
This allows us to perform the sum, and we arrive at
\ba
\lefteqn{\dot{\chi}_{S \lambda}(\theta,\phi) =  -i\dot{\phi}\cos\theta
  \left[J_{z'}, {\chi}_{S \lambda}(\theta,\phi)\right]} \nonumber\\
&+&{1\over 2}(v_\theta +iv_\phi)\sqrt{S(S+1)-\lambda(\lambda-1)}
        {\chi}_{S \lambda-1}(\theta,\phi) \nonumber\\
        &-&{1\over 2}(v_\theta -iv_\phi)\sqrt{S(S+1)-\lambda(\lambda+1)}
        {\chi}_{S \lambda+1}(\theta,\phi) \ . \nonumber\\
\label{spinco}
\ea
Here, we used $\left[J_{z'}, {\chi}_{S \lambda}(\theta,\phi)\right] =
\lambda {\chi}_{S \lambda}(\theta,\phi)$. To obtain the part relevant
for the covariant derivative we project the right-hand side of
Eq.~(\ref{spinco}) back onto ${\chi}_{S \lambda}(\theta,\phi)$. For a
general spin function $\eta(t)=\sum_\lambda \eta^\lambda(t) {\chi}_{S
  \lambda}(\theta,\phi)$ in the body-fixed system we thus have
\be
\label{covspin}
D_t \eta =\partial_t\eta  -i\dot{\phi}\cos\theta\left[J_{z'}, {\eta}
  \right] \ .
\ee
Had we written the vector $\mathbf{a}$ considered above in terms of
its spherical components we would have obtained the same result.
Applying the result~(\ref{covspin}) to the spinor functions
$\hat{\Psi}(\mathbf{x}')$ yields Eq.~(\ref{cov-diff}).

\section{Coset space}
\label{sec:coset}

We exploit the nonlinear realization of rotational invariance more
formally than done in the calculations of Apps.~\ref{trans} and
\ref{sec:cova}. Thereby we connect to previous EFTs on axially
deformed nuclei~\cite{papenbrock2011,papenbrock2014,coelloperez2015},
nuclei with tri-axial deformation~\cite{chen2017,chen2018,chen2020},
and magnets~\cite{roman1999,hofmann1999,baer2004,kampfer2005}. We
follow closely the original
papers~\cite{weinberg1968,coleman1969,callan1969}. For reviews of this
approach, and an exhibition for non-relativistic systems, we refer the
readers to Refs.~\cite{leutwyler1994,leutwyler1996,brauner2010} and
the textbook~\cite{weinbergbook}. In finite systems, one speaks of
emergent symmetry breaking~\cite{yannouleas2007} but the tools from
field theory can also be extended to this
case~\cite{gasser1988,papenbrock2011,papenbrock2014}. Not
surprisingly, the calculations in the present Section have much in
common with those in Appendices~\ref{rotsf} and \ref{rotbf}.

Three mutually orthogonal unit vectors $(|\mathbf{e}^\prime_x\rangle,
|\mathbf{e}^\prime_y\rangle, |\mathbf{e}^\prime_z \rangle)$ (the
``body-fixed system'') are linked to another three mutually orthogonal
unit vectors $(|{\bf e}_x\rangle$, $|{\bf e}_y \rangle$, $|{\bf e}_z
\rangle)$ (the ``space-fixed system'') by a rotation $g$ so that for
$k = x, y, z$ we have $|\mathbf{e}^\prime_k \rangle = g |{\bf e}_k
\rangle$. That can be written as $| \mathbf{e}^\prime_k \rangle =
\sum_j | {\bf e}_j \rangle \langle {\bf e}_j | g | {\bf e}_k \rangle$
$= \sum_j | {\bf e}_j \rangle g_{j k}$ where $g_{j k} = \langle {\bf
  e}_j | g | {\bf e}_k \rangle$ is the matrix representation of
$g$. The matrix $g_{j k}$ is real orthogonal, $g_{j k} = (g^{- 1})_{k
  j}$, hence $| {\bf e}_j \rangle = \sum_k g_{j k} |
\mathbf{e}^\prime_k \rangle$. Altogether,
\ba
| {\bf e}_j \rangle = \sum_k g_{j k} | \mathbf{e}^\prime_k \rangle \ , \
|\mathbf{e}^\prime_k \rangle = \sum_j (g^{- 1})_{k j} | {\bf e}_j \rangle \ .
\label{1}
\ea
For vectors we use small (capital) letters when they are written in
the space-fixed (the body-fixed) system, respectively. For a vector
${\bf a} = \sum_j a_j | {\bf e}_j \rangle$ we have ${\bf A} = \sum_j
A_j |\mathbf{e}^\prime_j \rangle$ where
\ba
A_j = \sum_k (g^{- 1})_{j k} a_k \ . 
\label{2}
\ea
The transformation $g$ is defined as
\ba
g(\theta, \phi) = \exp \{ - i \phi J_z \} \exp \{ - i \theta J_y \} \ .
\label{3}
\ea
It coincides with the transformation ${\cal R}(\phi, \theta, 0)$ in
Section~\ref{eerot}. The three generators $J_k$ of infinitesimal
rotations about the $k$-axes obey
\ba
[J_x, J_y] = i J_z \ {\rm (cyclic)} \ .
\label{4}
\ea
The matrix representation of the operators $J_x, J_y, J_z$ is
\ba
- i J_x &\to& \left( \begin{matrix} 0 & 0 & 0 \cr
  0 &  0  & - 1 \cr
  0 &  1 &  0 \cr
  \end{matrix}\right) \ , \nonumber \\ 
- i J_y &\to& \left( \begin{matrix} 0 & 0 & 1 \cr
  0 &  0  & 0 \cr
  -1 &  0 &  0 \cr
  \end{matrix} \right) \ , \nonumber \\ 
- i J_z &\to& \left( \begin{matrix} 0 & - 1 & 0 \cr
  1 &  0  & 0 \cr
  0 &  0 &  0 \cr
  \end{matrix} \right) \ .
\label{5}
\ea
The commutation relations~(\ref{4}) for the matrix representation are
verified using standard matrix algebra. The relations~(\ref{5}) imply
\ba
\exp \{ - i \phi J_z \} &\to& \left( \begin{matrix}
  \cos \phi & - \sin \phi & 0 \cr
  \sin \phi & \cos \phi & 0 \cr
  0 & 0 & 1 \cr
  \end{matrix} \right) \ , \nonumber \\
\exp \{ - i \theta J_y \} &\to& \left( \begin{matrix}
  \cos \theta & 0 & \sin \theta \cr
  0 & 1 & 0 \cr
  - \sin \theta & 0 & \cos \theta \cr
  \end{matrix} \right) \ ,
\label{6}
\ea
and, thus,
\ba
g &\to& \left( \begin{matrix}
  \cos \phi \cos \theta & - \sin \phi & \cos \phi \sin \theta \cr 
  \sin \phi \cos \theta &   \cos \phi & \sin \phi \sin \theta \cr
  - \sin \theta & 0 & \cos \theta \cr\end{matrix} \right) \ .
\label{7}
\ea
In the body-fixed system, we define a set of three operators
$\tilde{J}_k$, $k = x, y, z$. These have the same commutation
relations~(\ref{4}) as the operators $J_k$. Moreover, these operators
have, by definition, the same matrix representation~(\ref{5}) in the
basis $| \mathbf{e}^\prime_k \rangle$ as do the operators $J_k$ in the
basis $| {\bf e}_k \rangle$. Hence, with
\ba
g = \sum_\mu | \mathbf{e}^\prime_\mu \rangle \langle \mathbf{e}_\mu |
\ , \ g^{- 1} = \sum_\mu | \mathbf{e}_\mu \rangle \langle
\mathbf{e}^\prime_\mu |
\label{8}
\ea
we have for $k = x, y, z$
\ba
\tilde{J}_k = g J_k g^{- 1} \ , \ J_k = g^{- 1} \tilde{J}_k g \ .
\label{9}
\ea
The commutation relations for the operators $\tilde{J}_k$ differ in
sign from the anomalous commutators commonly used in the body-fixed
system. The reason is that the definition~(\ref{9}) employs the matrix
representation of $J_k$ on its right-hand side. Conventionally, when
using differential operators for $J_k$, these act also on the angles
in $g$ and one obtains additional transformation terms leading to
anomalous commutation relations. The operators $J_k$ and $\tilde{J}_k$
differ. That is seen by comparing the matrix representations in the
basis $| {\bf e}_l \rangle$,
\ba
\langle {\bf e}_l | \tilde{J}_k | {\bf e}_m \rangle = \sum_{n r} g_{l n} 
\langle {\bf e}_n | J_k | {\bf e}_r \rangle g_{m r} \ .
\label{10}
\ea
In analogy to Eq.~(\ref{3}) we define the operator
\ba
\tilde{g}(\theta, \phi) = \exp \{ - i \phi \tilde{J}_z \} \exp \{ - i
\theta \tilde{J}_y \} \ .
\label{11}
\ea
In the body-fixed system, the matrix elements of $\tilde{g}$ are given
by
\ba
\tilde{g}_{\mu \nu} = \langle \mathbf{e}^\prime_\mu | \tilde{g} |
\mathbf{e}^\prime_\nu
\rangle \ .
\label{12}
\ea
Eq.~(\ref{9}) implies that the matrix elements $g_{\mu \nu}$ of $g$
in the space-fixed system and $\tilde{g}_{\mu \nu}$ of $\tilde{g}$ in
the body-fixed system are equal,
\ba
g_{\mu \nu} = \tilde{g}_{\mu \nu} \ .
\label{13}
\ea
The equality of these two matrices implies that we may use either
form. If the matrix $g$ operates in the space-fixed system we use the
form $g_{\mu \nu}$, if it acts in the body-fixed system, we use the
form $\tilde{g}_{\mu \nu}$. If we employ an operator representation we
proceed likewise and use $g$ as defined in Eq.~(\ref{4}) in the
space-fixed system and $\tilde{g}$ as defined in Eq.~(\ref{11}) in the
body-fixed system.

Let the angles $\theta, \phi$ and, with these, the transformation $g$
be dependent upon time.  Let ${\bf A} = A_x | \mathbf{e}^\prime_x
\rangle + A_y | \mathbf{e}^\prime_y \rangle$ be a vector in the
tangential plane (i.e., perpendicular to $| \mathbf{e}^\prime_z
\rangle$) with time-dependent components $A_x(t), A_y(t)$. The time
derivative of ${\bf A}$, indicated by a dot, is
\ba
\dot{\bf A} = \dot{A}_x | \mathbf{e}^\prime_x \rangle + \dot{A}_y |
\mathbf{e}^\prime_y
\rangle + A_x | \dot{\mathbf{e}}^\prime_x \rangle + A_y |
\dot{\mathbf{e}}^\prime_y \rangle
\ .
\label{14}
\ea
We use $| \dot{\mathbf{e}}^\prime_k \rangle = \sum_j | {\bf e}_j
\rangle \dot{g}_{j k} = \sum_{j l} g_{j l} \dot{g}_{j k} |
\mathbf{e}^\prime_l \rangle$ and $ g_{j l} = g^{- 1}_{l j}$. Moreover,
from $({\rm d} / {\rm d} t) (g^{- 1} g) = 0$ it follows that the
matrix $(g^{- 1} \dot{g})_{k l}$ is antisymmetric. Thus,
\ba
| \dot{\mathbf{e}}^\prime_k \rangle = \sum_l (g^{- 1} \dot{g})_{l k} |
\mathbf{e}^\prime_l
\rangle = - \sum_l (g^{- 1} \dot{g})_{k l} | \mathbf{e}^\prime_l \rangle \ .
\label{15}
\ea
Explicit calculation shows that
\ba
g^{- 1} \dot{g} \to \left( \begin{matrix}
  0 & - \dot{\phi} \cos \theta & \dot{\theta} \cr
  \dot{\phi} \cos \theta & 0 & \dot{\phi} \sin \theta \cr
  - \dot{\theta} & - \dot{\phi} \sin \theta & 0 \cr\end{matrix} \right) \ .
\label{16}
\ea
Combining Eqs.~(\ref{9}) to (\ref{11}) we obtain
\ba
\dot{\bf A} &=& \dot{A}_x | \mathbf{e}^\prime_x \rangle + \dot{A}_y |
\mathbf{e}^\prime_y
\rangle + A_x \dot{\phi} \cos \theta | \mathbf{e}^\prime_y \rangle - A_x
\dot{\theta} | \mathbf{e}^\prime_z \rangle \nonumber \\
&& \qquad - A_y \dot{\phi} \cos \theta | \mathbf{e}^\prime_x \rangle - A_y
\dot{\phi} \sin \theta | \mathbf{e}^\prime_z \rangle \ .
\label{17}
\ea
This is Eq.~(\ref{cov-diff1}). The covariant derivative of ${\bf A}$
is defined as the projection of $\dot{\bf A}$ onto the tangential
plane,
\ba
D_t {\bf A} = ({\dot A}_x - A_y \dot{\phi} \cos \theta) | \mathbf{e}^\prime_x
\rangle + ({\dot A}_y + A_x \dot{\phi} \cos \theta) | \mathbf{e}^\prime_y
\rangle \ .\nonumber\\
\label{18}
\ea
Using the fact that in the basis $| \mathbf{e}^\prime_k \rangle$ the
operators $\tilde{J}_k$ have the matrix representation~(\ref{5}), we
write Eq.~(\ref{18}) in the form
\ba
i D_t {\bf A} = (i \partial_t + \dot{\phi} \cos \theta \tilde{J}_z)
{\bf A} \ .
\label{19}
\ea
That agrees with Eq.~(\ref{cov-diff}). The partial derivative acts
only on the components $(A_x, A_y)$ of ${\bf A}$. The additional term
accounts for a rotation around the $z'$-axis by the angle $\dot{\phi}
\cos \theta$. That is the hallmark of a covariant derivative.

Given two vectors ${\bf a} = \sum_j a_j(t)$ $| {\bf e}_j \rangle$ and
${\bf b} = \sum_j b_j(t) | {\bf e}_j \rangle$ in the space-fixed
system with time-dependent coefficients $a_j(t), b_j(t)$, we
transcribe the inner product of ${\bf b}$ and of the time derivative
$\dot{\bf a}$ of ${\bf a}$, i.e., the expression $\sum_j b_j
\dot{a}_j$, into the body-fixed system. As mentioned earlier we
distinguish the system-dependent representations of the two vectors by
writing ${\bf a} \to {\bf A} = \sum_j A_j | \mathbf{e}^\prime_j
\rangle$ and ${\bf b} \to {\bf B} = \sum_j B_j | \mathbf{e}^\prime_j
\rangle$. To focus attention on the covariant derivative we put $A_z =
0 = B_z$. Then both ${\bf A} = A_x | \mathbf{e}^\prime_x \rangle + A_y
| \mathbf{e}^\prime_y \rangle$ and ${\bf B} = B_x |
\mathbf{e}^\prime_x \rangle + B_y | \mathbf{e}^\prime_y \rangle$ are
tangential vectors in the body-fixed system. From Eq.~(\ref{2}) we
have $a_j = \sum_k g_{j k} A_k$, $b_j = \sum_k g_{j k} B_k$ and, thus,
\ba
   {\bf b} \dot{\bf a} &=& \sum_j b_j \dot{a}_j \nonumber\\
   &=& \sum_{j k l} B_k g_{j k}
   \frac{\rm d}{{\rm d} t} (g_{j l} A_l) \nonumber\\
   &=& \sum_k B_k \dot{A}_k +
   \sum_{k l} B_k A_l (g^{-1} \dot{g})_{k l} \nonumber \\
   &=& B_x (\dot{A}_x - \dot{\phi} \cos \theta A_y) + B_y (\dot{A_y} +
   \dot{\phi} \cos \theta A_x) \ ,\nonumber\\
\label{20}
\ea
or, using the definition~(\ref{19}),
\ba
{\bf b} \dot{\bf a} = {\bf B} D_t {\bf A} \ .
\label{21}
\ea
Eq.~(\ref{21}) gives the rule for transcribing time derivatives of
vectors into the body-fixed system. It applies provided in the
body-fixed system, the vectors are tangential.

We define an infinitesimal transformation $r$ in the space-fixed
system and another infinitesimal transformation $\tilde{r}$ in the
body-fixed system. Both are defined in terms of the augmented rotation
$g(\theta + \delta \theta, \phi + \delta \phi) \exp \{ - i J_z \gamma
\}$. Here $\delta \theta, \delta \phi, \gamma$ are infinitesimal. That
changes $g \to g + \delta g$. In the space-fixed system we consider
the infinitesimal transformation $\delta g$ acting on the vectors $|
{\bf e}_j \rangle$, keeping the vectors $| \mathbf{e}^\prime_k
\rangle$ fixed.  Eqs.~(\ref{1}) give
\ba
| \delta {\bf e}_j \rangle &=& \sum_k (\delta g)_{j k} |
\mathbf{e}^\prime_k \rangle \nonumber\\
&=& \sum_{k l} (\delta g)_{j k} g_{l k} | {\bf e}_l \rangle \nonumber\\
&=&
\sum_l (\delta g g^{- 1})_{j l} | {\bf e}_l \rangle \nonumber\\
&=& \sum_l r_{j l} |
    {\bf e}_l \rangle \ .
\label{22}
\ea
The last relation defines $r$. Explicit calculation shows that
\ba
r &=& \delta g g^{- 1} \nonumber\\
&=& \delta \phi (- i J_z) + \delta \theta \cos \phi
(- i J_y) \nonumber\\
&&- \delta \theta \sin \phi (- i J_x) + \gamma \cos \theta
(- i J_z) \nonumber \\
&& + \gamma \sin \theta \cos \phi (- i J_x) + \gamma \sin \theta
\sin \phi (- i J_y) \ .
\label{23}
\ea
A general infinitesimal transformation in the space-fixed system can
be written in terms of infinitesimal angles $\delta \alpha_k$ as
\ba
r = \sum_k \delta \alpha_k (- i J_k) \ .
\label{24}
\ea
Equating that with $r$ as given in Eq.~(\ref{23}) we obtain a linear
relation between the infinitesimal angles $\delta \theta$, $\delta
\phi$, $\gamma$ and the angles $\delta \alpha_k$. It reads
\ba
\left(
\begin{array}{c} \delta \alpha_x \\ \delta \alpha_y \\ \delta \alpha_z
\end{array} \right) = \left( \begin{array}{ccc}
  - \sin \phi & 0 & \sin \theta \cos \phi \\ \cos \phi &  0 & \sin
  \theta \sin \phi \\ 0 & 1 & \cos \theta \end{array} \right)
\left( \begin{array}{c} \delta \theta \\ \delta \phi \\ \gamma
\end{array} \right) \ .\nonumber\\
\label{25}
\ea
The inverse relation is
\ba
\left(
\begin{array}{c} \delta \theta \\ \delta \phi \\ \gamma \end{array}
\right) = \left( \begin{array}{ccc}
- \sin \phi & \cos \phi & 0 \\ - \cos \phi \cot \theta & - \sin
\phi \cot \theta & 1 \\ {\cos \phi \over \sin \theta} & {\sin \phi
  \over \sin \theta} & 0 \end{array} \right) \left( \begin{array}{c}
  \delta \alpha_x \\ \delta \alpha_y \\ \delta \alpha_z \end{array}
\right) \ . \nonumber\\
\label{26}
\ea
Identifying $\gamma$ with $\delta \omega$, we see that this agrees
with Eqs.~(\ref{trafo_theta_phi}) and (\ref{gamma}). In the body-fixed
system we consider the infinitesimal transformation $(\delta g)^{- 1}$
acting on the vectors $| \mathbf{e}^\prime_k \rangle$, keeping the
vectors $| {\bf e}_j \rangle$ fixed.  Equations~(\ref{1}) give
\ba
| \delta\mathbf{e}^\prime_k \rangle &=& \sum_j (\delta{g}^{- 1})_{k j} |
{\bf e}_j \rangle \nonumber\\
&=& \sum_{j l} (\delta{g}^{- 1})_{k j} g_{j l} | \mathbf{e}^\prime_l \rangle
\nonumber\\
&=& \sum_l [(\delta{g})^{- 1} g]_{k l} | \mathbf{e}^\prime_l \rangle
\nonumber \\
&=& \sum_l \tilde{r}_{k l} | \mathbf{e}^\prime_l \rangle \ .
\label{27}
\ea
The last relation defines $\tilde{r}$. Eqs.~(\ref{27}) show that
$\tilde{r}$ acts in the body-fixed system. We use the arguments below
Eq.~(\ref{13}) to express $\tilde{r}$ in terms of the operator
$\tilde{g}$ defined in Eq.~(\ref{11}). Explicit calculation shows that
\ba
\tilde{r} &=& (\delta \tilde{g})^{- 1} \tilde{g} \nonumber\\
&=& - \delta \phi \cos
\theta (- i \tilde{J}_z) - \delta \theta (- i \tilde{J}_y) \nonumber\\
&&+ \delta \phi
\sin \theta (- i \tilde{J}_x) - \gamma (- i \tilde{J}_z) \ .
\label{28}
\ea
Since $\delta (\tilde{g}^{- 1} \tilde{g}) = 0$ we have $(\delta
\tilde{g}^{- 1}) \tilde{g} = - \tilde{g}^{- 1} \delta \tilde{g}$. The
last relation shows that the three infinitesimal angles $\delta
\theta$, $\delta \phi$, $\gamma$ all carry negative signs. That is
because the infinitesimal transformation $\delta \tilde{g}^{- 1}$
acts conversely to the infinitesimal transformation $\delta g$. A
general infinitesimal transformation in the body-fixed system can be
written in terms of infinitesimal angles $\delta \tilde{\alpha}_k$ as
\ba
\tilde{r} = \sum_k \delta \tilde{\alpha}_k (- i \tilde{J}_k) \ .
\label{29}
\ea
Since in Eq.~(\ref{28}) $\tilde{r}$ acts conversely to $r$ we equate
expression~(\ref{28}) not with expression~(\ref{29}) but with the
converse of expression~(\ref{29}), obtained by the replacements
$\delta \tilde{\alpha}_k \to - \delta \tilde{\alpha}_k$ for all
$k$. That gives
\ba
\left( \begin{array}{c} \delta \tilde{\alpha}_x \\ \delta
  \tilde{\alpha}_y \\ \delta \tilde{\alpha}_z \end{array} \right) =
\left( \begin{array}{ccc} 0 & - \sin \theta & 0 \\ 1 & 0 & 0 \\
  0 & \cos \theta & 1 \end{array} \right)
\left( \begin{array}{c} \delta \theta \\ \delta \phi \\ \gamma
\end{array} \right) \ .
\label{30}
\ea
The inverse relation is
\ba
\left( \begin{array}{c} \delta \theta \\ \delta \phi \\ \gamma
\end{array} \right) = \left( \begin{array}{ccc} 0 & 1 & 0 \\
  - {1 \over \sin \theta} & 0 & 0 \\ \cot \theta & 0 & 1 \end{array}
\right) \left( \begin{array}{c} \delta \tilde{\alpha}_x \\ \delta
  \tilde{\alpha}_y \\ \delta \tilde{\alpha}_z \end{array} \right) \ .
\label{31}
\ea
Identifying $\gamma$ with $\delta \omega$, we see that this agrees
with Eqs.~(\ref{bf-trafo_theta_phi}) and (\ref{bf-gamma}). 

The commutation relations~(\ref{4}) imply that
  under the infinitesimal rotation $1 + \delta \alpha_z (- i J_z)$,
  the operator $(- i J_x)$ is mapped onto $[1 + \delta \alpha_z (- i
    J_z)] (- i J_x) [1 - \delta \alpha_z (- i J_z)] = (- i J_x) +
  \delta \alpha_z (- i J_y)$, and correspondingly for the other
  components. That shows that under a rotation, the three operators
  $(- i J_x, - i J_y, - i J_z)$ transform like the three unit vectors
  $({\bf e}_x, {\bf e}_y, {\bf e}_z)$ of a three-dimensional linear
  space. That suggests that $r$ in Eq.~(\ref{24}) and $\tilde{r}$ in
  Eq.~(\ref{29}) represent the same vector written, respectively, in
  the space-fixed and in the body-fixed coordinate system. For that to
  be true the three infinitesimal angles $(\delta \alpha_x, \delta
  \alpha_y, \delta \alpha_z)$ of $r$ and $(\delta \alpha_x, \delta
  \alpha_y, \delta \alpha_z)$ of $\tilde{r}$ must be connected as in
  Eq.~(\ref{2}),
\ba
\delta \alpha_k = \sum_l g_{k l} \delta \tilde{\alpha}_l \ .
\label{47a}
\ea
Combining Eqs.~(\ref{25}) and (\ref{31}) yields
\ba
&& \left( \begin{array}{c} \delta \alpha_x \\ \delta \alpha_y \\ \delta
  \alpha_z \end{array} \right) \nonumber  \\
&& = \left( \begin{array}{ccc} \cos \theta
  \cos \phi & - \sin \phi & \sin \theta \cos \phi \\
  \cos \theta \sin \phi & \cos \phi & \sin \theta \sin \phi \\
  - \sin \theta & 0 & \cos \theta \end{array}
\right) \left( \begin{array}{c} \delta \tilde{\alpha}_x \\ \delta
  \tilde{\alpha}_y \\ \delta \tilde{\alpha}_z \end{array} \right) \ .\nonumber\\
\label{47b}
\ea
Equation~(\ref{7}) shows that Eq.~(\ref{47b}) indeed equals
Eq.~(\ref{47a}), confirming the vector character of $r$. Applying that
to Noether's theorem in App.~\ref{sec:emmy} we see that $I_k$ and
$I_{k'}$ are indeed the components of the same vector written
respectively, in the space-fixed and in the body-fixed coordinate
system.

It is straightforward to extend these arguments from vectors,
i.e. spherical tensors of rank three, to spherical tensors of
arbitrary rank. Then, the concrete representations of the rotation
matrices $g$ and $\tilde{g}$ are given by Wigner $D$ matrices, while
all algebraic relationships derived above remain unchanged.

\section{Gauge potentials}
\label{sec:gauge1}

We demonstrate how gauge potentials arise in an adiabatic approach and
we discuss gauge transformations and their relation to rotations.

\subsection{Gauge potentials from an adiabatic approach}
\label{sec:berry}

The appearance of the non-Abelian gauge potential~(\ref{Aabel}) can be
understood also in an adiabatic
approach~\cite{kuratsuji1985,wilczek1989}. If the nucleon's degrees of
freedom are much faster than those of the rotor, the eigenstates of
the fermion Hamiltonian $H_\Psi$ follow the rotor's axial symmetry
instantaneously, independently of any details of the fermion-rotor
interaction. For simplicity we consider only the fermion spin function
$\chi_{Sm}$ with half-integer spin $S$ and projection $m$ onto the
space-fixed $z$ axis. As the fermion is fast, it's spin is in an
eigenstate with respect to projection onto the rotor's symmetry axis,
i.e. the helicity spin states $\chi_{S\lambda}(\theta,\phi)$ from
Eq.~(\ref{helicity}) span, for fixed projection $\lambda$, a basis of
the instantaneous fermion eigenstates. They fulfill
\be
(\mathbf{e}_r\cdot\mathbf{S}) \, \chi_{S \lambda}(\theta,\phi) = \lambda
\chi_{S \lambda}(\theta,\phi) \ .
\ee
Due to Kramers' degeneracy, the spin states $\chi_{S {\pm
    \lambda}}(\theta,\phi)$ are degenerate. In the adiabatic
approximation, one evaluates the Hamiltonian of the fermion-plus-rotor
system
\be
H = H_{\Psi} - {1\over 2C_0} \nabla_\Omega^2  \ ,
\ee
in these eigenstates to get the effective Hamiltonian matrix (see,
e.g., Berry's overview in Ref.~\cite{wilczek1989})
\ba
H_{S'S} &\equiv& \chi^\dagger_{S'\lambda}(\theta,\phi) H\chi_{S \lambda}
(\theta,\phi) \nonumber\\
&=&  {1\over 2C_0} \left(-i\delta_{S'S'}\nabla_\Omega - \mathbf{A}_{S' S}
\right)^2 \nonumber\\
&&+ \chi^\dagger_{S\lambda}(\theta,\phi)H_{\Psi} \chi_{S\lambda}(\theta,\phi)
\ .
\ea
Here the vector gauge potential is the matrix
\ba
\mathbf{A}_{S' S}  &\equiv& i \chi^\dagger_{S'\lambda}(\theta,\phi)\nabla_\Omega
\chi_{S,\lambda}(\theta,\phi) \ .
\ea
Using properties of Wigner $D$ functions~\cite{varshalovich1988} and a
summation formula from Ref.~\cite{lai1998}, one finds
\ba
\label{AS'S}
\mathbf{A}_{S'S} &=& \delta_{S'S}\lambda \cot\theta \mathbf{e}_\phi \ .
\ea
In our case, the projection $\lambda$ is obtained by application of
the operator $\hat{K}_{z'}$, and we thus find that the gauge potential
$\mathbf{A}=\mathbf{e}_\phi \cot\theta \hat{K}_{z'}$ enters. This is
Eq.~(\ref{Aabel}).

\subsection{Gauge transformations}
\label{sec:gaugetrafo}

Let us also explore gauge transformations. Our definition of the
body-fixed coordinate system~(\ref{eprime}) is convenient [because the
  basis vectors $\mathbf{e}_\theta$ and $\mathbf{e}_\phi$ are tangent
  vectors of the  lines parameterized by the spherical coordinates
  $(\theta,\phi)$] but otherwise arbitrary. Any rotation of these
basis vectors around the $\mathbf{e}_r$ axis would have been equally
valid, i.e. the vectors
\ba
\mathbf{e}^\prime_1&\equiv&\cos\gamma(\theta,\phi)\mathbf{e}_\theta +
\sin\gamma(\theta,\phi)\mathbf{e}_\phi \ , \nonumber\\
\mathbf{e}^\prime_2&\equiv&-\sin\gamma(\theta,\phi)\mathbf{e}_\theta +
\cos\gamma(\theta,\phi)\mathbf{e}_\phi \ ,
\ea
and $\mathbf{e}_r$ span a right-handed body-fixed coordinate
system. Here, $\gamma(\theta,\phi)$ is a smooth function over the
sphere. Let us repeat the computations made in the previous Subsection
for this body-fixed system. The helicity basis functions for the
fermion become
\be
\tilde{\chi}_{S \lambda}(\theta,\phi) = \sum_m D^S_{m \lambda}(\phi,
\theta,\gamma) \chi_{Sm} \ . 
\ee
Here, and in what follows we suppress the dependence of $\gamma$ on
the angles $(\theta,\phi)$. The gauge potential is
\ba
\tilde{\mathbf{A}}_{S'S}  &\equiv& i \tilde{\chi}^\dagger_{S'\lambda}
(\theta,\phi)\nabla_\Omega \tilde{\chi}_{S,\lambda}(\theta,\phi)
\nonumber\\
&=& i\delta_{S'S}\sum_m \left[D^S_{m \lambda}(\phi,\theta,\gamma)
  \right]^* \nabla_\Omega D^S_{m\lambda}(\phi,\theta,\gamma)
\nonumber\\
&=& i\delta_{S'S}\sum_m d^S_{m\lambda}(\theta)\bigg[\left({-im\over
    \sin\theta}\mathbf{e}_\phi -i\lambda \nabla_\Omega \gamma \right)
  d_{m\lambda}^S(\theta) \nonumber\\
  &&-{1\over 2}\sqrt{S(S+1)-m(m-1)} d_{m-1 \lambda}^S(\theta)
  \mathbf{e}_\theta\nonumber\\
  &&+{1\over 2}\sqrt{S(S+1)-m(m+1)} d_{m+1 \lambda}^S(\theta)
  \mathbf{e}_\theta\bigg]\nonumber\\
&=& \delta_{S'S} \lambda \left[\cot\theta \mathbf{e}_\phi +
  \nabla_\Omega \gamma(\theta,\phi) \right] \ .
\ea
We have used results from chapter 4.9 of
Ref.~\cite{varshalovich1988}. The sums over the last two terms cancel
each other, and we used $\sum_m m[d_{m\lambda}(\theta)]^2
=\lambda\cos\theta$ and $\sum_m [d_{m\lambda}(\theta)]^2 =1$ from
Ref.~\cite{lai1998}.

The vector potentials $\tilde{\mathbf{A}}_{S'S}$ and
$\mathbf{A}_{S'S}$ differ by a gauge transformation, which is
generated by the arbitrary angle $\gamma(\theta,\phi)$. For
$\gamma(\theta,\phi) = \pm\phi$, for instance, one obtains the gauge
potentials $\tilde{\mathbf{A}}=\lambda{\cos\theta\pm
  1\over\sin\theta}\mathbf{e}_\phi$ by \textcite{wu1976}. Another
interesting choice is $\gamma(\theta,\phi) = -\phi\cos\theta$, because
it generates the non-singular gauge potential
$\tilde{\mathbf{A}}=\lambda\phi\sin\theta\mathbf{e}_\theta$.  Our
gauge potential~(\ref{Aabel}) is singular at both poles, and the rotor
eigenfunctions are Wigner $D$ functions. As pointed out in
Ref.~\cite{dray1986}, the different gauge potentials correspond to
different conventions regarding the third argument of the Wigner $D$
function, i.e. to different conventions regarding rotations of the
body-fixed coordinate system around its $z'$ axis. In other words, the
wave function $D_{M K}^I(\phi,\theta,0)$ we used in the main text is
replaced by $D_{M K}^I[\phi,\theta,\gamma(\theta,\phi)]$ when a gauge
transformation is made.

The arguments of this Subsection show that the freedom of choice of
the intrinsic coordinate system introduces a gauge freedom in the
dynamics of the collective rotational degrees of
freedom~\cite{littlejohn1997}. The general Abelian gauge potential is
\be
\tilde{\mathbf{A}}_{\rm a}(\theta,\phi) = \hat{K}_{z'}\left[
  \mathbf{e}_\phi \cot\theta+ \nabla_\Omega \gamma(\theta,\phi)
  \right] \ .
\ee

We extend the discussion to the non-Abelian gauge field. In its
manifestly gauge-invariant form it reads
\ba
\label{Anon2}
\tilde{\mathbf{A}}_{\rm n}(\theta,\phi) &=& g \mathbf{e}_r\times
\hat{\mathbf{K}} \nonumber\\
&=& g \left(\hat{K}_1\mathbf{e}^\prime_2 -  \hat{K}_2\mathbf{e}^\prime_1
\right) \nonumber\\
&=& g\Big[ \left(\hat{K}_1\cos\gamma(\theta,\phi) -\hat{K}_2
  \sin\gamma(\theta,\phi)\right)\mathbf{e}_\phi \nonumber\\
  &&-  \left(\hat{K}_1\sin\gamma(\theta,\phi) +\hat{K}_2
  \cos\gamma(\theta,\phi) \right)\mathbf{e}_\theta \Big] \ .
\nonumber\\
\ea
We have employed the operators $\hat{K}_i\equiv
\mathbf{e}^\prime_i\cdot\hat{\mathbf{K}}$ for $i=1,2$ in the
body-fixed frame. In the last line we have expressed the result in the
coordinate system of the rotor's angular velocity. We have used that
the spin operators $\hat{K}_1$, $\hat{K}_2$, and $\hat{K}_{z'}$
fulfill the canonical commutation relations. Comparison with
Eq.~(\ref{Anon}) meets our expectations, because the spin operators
are simply rotated by $\gamma$ around the body-fixed $z'$ axis.

The magnetic field of the total gauge potential
$\tilde{\mathbf{A}}_{\rm tot} = \tilde{\mathbf{A}}_{\rm a} +
\tilde{\mathbf{A}}_{\rm n}$ is
\ba
\tilde{\mathbf{B}}_{\rm tot} &=& \nabla_\Omega\times
\tilde{\mathbf{A}}_{\rm tot} -i \tilde{\mathbf{A}}_{\rm tot}\times
\tilde{\mathbf{A}}_{\rm tot} \nonumber\\
&=& (g^2-1)\hat{K}_{z'} \mathbf{e}_r \ , 
\ea
in agreement with Eq.~(\ref{Btot}). It does not depend on
$\gamma(\theta,\phi)$ and is, thus, gauge invariant.

We turn this argument around and study the effect of rotations on the
gauge potential. While the non-Abelian part~(\ref{Anon2}) of the gauge
potential is manifestly invariant under rotations, this is not so for
the Abelian part~(\ref{Aabel}). Under a rotation we have
\ba
\lefteqn{{\cal R} \mathbf{A}_{\rm a}(\theta,\phi) =}\nonumber\\
&& \hat{K}_{z'} \cot\theta \left[\mathbf{e}_\phi(\phi+\delta\phi,
  \theta+\delta\theta)  -\delta\omega\mathbf{e}_\theta(\phi+\delta
  \phi,\theta+\delta\theta) \right] \ .\nonumber\\
\ea
Here, the differential $\delta\omega$ is given in Eq.~(\ref{gamma})
for rotations around the space-fixed axes.  This rotated gauge
potential has to be compared with the gauge potential
\ba
\lefteqn{\mathbf{A}_{\rm a}(\theta+\delta\theta,\phi+\delta\phi) =}
\nonumber\\
&& \hat{K}_{z'} \cot(\theta+\delta\theta) \mathbf{e}_\phi(\phi+\delta
\phi,\theta+\delta\theta)  
\ea
at the point $(\phi+\delta\phi,\theta+\delta\theta)$. Here, the
differential $\delta\theta$ is taken from
Eq.~(\ref{trafo_theta_phi}). The difference
\ba
\delta\mathbf{A} &=&{\cal R} \mathbf{A}_{\rm a}(\theta)  -
\mathbf{A}_{\rm a}(\theta+\delta\theta) \nonumber\\&=&-\delta\omega
\cot\theta   \hat{K}_{z'} \mathbf{e}_\theta(\phi+\delta\phi,\theta
+\delta\theta) \nonumber\\
&&+\delta\theta {\hat{K}_{z'}\over\sin^2\theta}\mathbf{e}_\phi(\phi
+\delta\phi,\theta+\delta\theta) 
\ea
can be written as $\delta\mathbf{A} = \nabla_\Omega
\hat{K}_{z'}\delta\omega$ when employing the expressions
~(\ref{trafo_theta_phi}) and (\ref{gamma}). Thus, after a rotation the
gauge potential can be brought back into its original
form~(\ref{Atot}) by performing a gauge
transformation~\cite{fierz1944}.

\section{Supplements to Section~(\ref{lagr})}
\label{sec:suppl}

To compute the contribution of the term linear in $g$ of the
Hamiltonian~(\ref{Hfinal}) for $K = 1 / 2$ states we introduce
spherical components
\ba
I_{\pm 1} &\equiv& \mp {1\over \sqrt{2}}\left(I_{x'}\pm iI_{y'}\right)
\nonumber\\
&=& {i\over \sqrt{2}}\left(i\partial_\theta \mp{1\over\sin\theta}
\partial_\phi \pm i\hat{K}_{z'}\cot\theta\right)
\ea
and 
\ba
\hat{K}_{\pm 1} &\equiv& \mp {1\over \sqrt{2}}\left(\hat{K}_{x'}\pm i
\hat{K}_{y'}\right) \ .
\ea
We write the term as
\be
\label{gCoriolis}
      {g\over C_0}\left(I_{x'}\hat{K}_{x'} +I_{y'}\hat{K}_{y'}\right)  =
      - {g\over C_0}\left(I_{-1}\hat{K}_{+1} +I_{+1}\hat{K}_{-1}\right) \ . 
\ee
Using the properties of the raising and lowering operators (see
chapters 3.1 and 4.2 in Ref.~\cite{varshalovich1988}) we find
\be
\hat{K}_{\pm 1} \left|{\mp {1\over 2}}\right\rangle = \mp {1\over\sqrt{2}}
\left|{\pm {1\over 2}}\right\rangle \ , 
\ee
and
\ba
\label{IonD}
\lefteqn{I_{\pm 1}D_{M, M'}^I(\phi,\theta,0) =}  \nonumber\\
&&\pm\sqrt{I(I+1)-M'(M'\pm 1)\over 2} D_{M, M'\pm 1}^I(\phi,\theta,0)\ .
\nonumber\\
\ea
Thus,
\begin{align}
  &\left(I_{-1}\hat{K}_{+1} +I_{+1}\hat{K}_{-1}\right) D^I_{M, {\pm {1\over 2}}}
  (\phi,\theta,0) \left|{\mp {1\over 2}}\right\rangle =
  \nonumber\\
  &\left(I+{1\over 2}\right)D^I_{M, {\mp {1\over2}}}(\phi,\theta,0)
  \left|{\pm {1\over 2}}\right\rangle \ .
\end{align}
Inspection shows that that the linear combinations 
\be
D^I_{M, {-{1\over2}}}(\phi,\theta,0) \left|{{1\over 2}}\right\rangle
+ (-1)^{I+{1\over 2}} D^I_{M, {1\over2}}(\phi,\theta,0) \left|{-{1\over 2}}
\right\rangle
\ee
are solutions of the Hamiltonian~(\ref{Hfinal}) for $K=1/2$. The phase
$(-1)^{I+{1\over 2}}$ results from the requirement that the odd-mass
nucleus is invariant under rotations by $\pi$ around any axis
perpendicular to the symmetry axis. Hence, the contribution from the
term proportional to $g$ in the Hamiltonian~(\ref{Hfinal}) becomes
\be
\label{deltaEg}
\Delta E(g) = -{g\over C_0} \delta_{|K|}^{1\over 2}(-1)^{I+{1\over 2}}
\left(I+{1\over 2}\right) \ .
\ee
That yields Eq.~(\ref{Eresult}).

We next compute the matrix elements of the $g$-dependent terms of the
Hamiltonian~(\ref{Hfinal}) for two close-lying band heads. Using the
normalization to $4\pi/(2I+1)$ of the squared Wigner function, see
Chapter 4.11 of Ref.~\cite{varshalovich1988}, we find
\begin{align}
  \int\limits_0^{2\pi}{\rm d}\phi &\int\limits_{-1}^{1}{\rm d}\cos\theta
  \langle K|\left[D^I_{M,-K}(\phi,\theta,0)\right]^* I_{+1}\hat{K}_{-1}
  \nonumber\\
&D^I_{M,-K-1}(\phi,\theta,0)|K{+1}\rangle \nonumber\\
=& \langle K| \hat{K}_{-1}|K{+1}\rangle  \nonumber\\
\times& \int\limits_0^{2\pi}{\rm d}\phi \int\limits_{-1}^{1}{\rm d}
\cos\theta D^{I*}_{M,-K}(\phi,\theta,0) I_{+1}D^I_{M,-K-1}(\phi,\theta,0)
\nonumber\\
=&   \frac{4\pi}{2I+1} \sqrt{I(I+1)-K(K +1)\over 2}  \langle K|
\hat{K}_{-1}|K{+1}\rangle\ .
\end{align}
We have used Eq.~(\ref{IonD}). The other relevant matrix element is
\begin{align}
  \int\limits_0^{2\pi}{\rm d}\phi &\int\limits_{-1}^{1}{\rm d}\cos\theta
  \langle {-K}|\left[D^I_{M,K}(\phi,\theta,0)\right]^* I_{-1}\hat{K}_{+1}
  \nonumber\\
&D^I_{M,K+1}(\phi,\theta,0)|{-K}{-1}\rangle \nonumber\\
=& \langle {-K}| \hat{K}_{+1}|{-K}{-1}\rangle  \nonumber\\
\times& \int\limits_0^{2\pi}{\rm d}\phi \int\limits_{-1}^{1}{\rm d}
\cos\theta D^{I*}_{M,K}(\phi,\theta,0) I_{-1}D^I_{M,K+1}(\phi,\theta,0)
\nonumber\\
=&  \frac{-4\pi}{2I+1} \sqrt{I(I+1)-K(K +1)\over 2}  \langle {-K}|
\hat{K}_{+1}|{-K}{-1}\rangle\ .
\end{align}
Time-reversal invariance relates both matrix elements. Denoting the
time-reversal operator by ${\cal T}$ we have
\ba
\langle -K| \hat{K}_{+1}|{-K}{-1}\rangle &=& \langle K|{\cal T}^\dagger
\hat{K}_{+1} {\cal T}|K{+1}\rangle \nonumber\\
&=& -\langle K|\hat{K}_{-1} |K{+1}\rangle \ .
\ea
The interaction is characterized by a single parameter. For a given
potential $V$, the relevant matrix element can be calculated by
expanding the axially symmetric eigenstates in terms of spherical
basis functions. In our approach, $\langle K|
\hat{K}_{-1}|K{+1}\rangle$ is a low-energy constant and needs to be
adjusted to data.

The next-to-leading-order correction of the Hamiltonian is
\ba
&& H_{\rm NLO} = -{1\over 2}\left(p_\theta+g\hat{K}_{y'},
p_\phi-\cos\theta \hat{K}_{z'}-g\sin\theta \hat{K}_{x'}\right)
\nonumber\\
&& \qquad \times \hat{M}_{\rm LO}^{-1}\hat{M}_{\rm
  NLO}\hat{M}_{\rm LO}^{-1} \left(\begin{array}{c}
  p_\theta+g\hat{K}_{y'}\\ p_\phi -\cos\theta \hat{K}_{z'}-g\sin\theta
  \hat{K}_{x'}\end{array}\right) \ . \nonumber\\ \ea
Here, the ``mass" matrices 
\ba
\hat{M}_{\rm LO} &=& {1\over C_0}\left[\begin{array}{cc}
1 & 0\\
0 & \sin^2\theta\end{array}\right] 
\ea
and
\ba
\hat{M}_{\rm NLO} &=& \left(g_a\left(\hat{K}_{x'}^2+\hat{K}_{y'}^2\right)
+g_b\hat{K}_{z'}^2 \right) \left[\begin{array}{cc}
1& 0   \\
0 &  \sin^2\theta\end{array}\right] \nonumber\\
&+&g_c \left[\begin{array}{cc}
    \hat{K}_{x'}^2 & \hat{K}_{x'}\hat{K}_{y'}\sin\theta\\
\hat{K}_{y'}\hat{K}_{x'}\sin\theta & \hat{K}_{y'}^2\sin^2\theta\end{array}
\right] 
\ea
enter the perturbative inversion of the mass matrix
\ba
\hat{M} = \hat{M}_{\rm LO} + \hat{M}_{\rm NLO}
\ea
via
\be
\hat{M}^{-1}\approx \hat{M}_{\rm LO}^{-1} - \hat{M}_{\rm LO}^{-1}
\hat{M}_{\rm NLO}\hat{M}_{\rm LO}^{-1} \ .
\ee
The resulting Hamiltonian is written as in Eq.~(\ref{Hnlo}). Using
Eq.~(\ref{Irot}) we replace the canonical momenta by angular momentum
components,
\ba
p_\theta &=& I_{y'} \ , \nonumber\\
{p_\phi\over \sin\theta} &=& \hat{K}_{z'}\cot\theta -I_{x'} \ , 
\ea
and find
\ba
\hat{C} &\equiv& -{\left(g_a+{g_c\over 2}\right)\left( \hat{K}_{x'}^2
  +\hat{K}_{y'}^2 \right)+g_b\hat{K}_{z'}^2 \over C_0} \left[
  \begin{array}{cc} 1 & 0\\
0 &1\end{array}\right] \ , \nonumber\\
\hat{G}&\equiv& -{g_c\over C_0}\left[\begin{array}{cc}
    {1\over 2}\left(\hat{K}_{x'}^2 -\hat{K}_{y'}^2\right)& \hat{K}_{x'}
    \hat{K}_{y'} \\
    \hat{K}_{y'}\hat{K}_{x'}  & {1\over 2}\left(\hat{K}_{y'}^2-\hat{K}_{x'}^2
    \right) \end{array}\right] \ , \nonumber\\
\ea
and
\ba
\mathbf{N}&\equiv& 
\left(\begin{array}{c}
I_{y'}\\
I_{x'} \end{array}\right)
+g\left(\begin{array}{c}
\hat{K}_{y'}\\
\hat{K}_{x'} \end{array}\right)
 \ .
\ea
With a view on Eq.~(\ref{Hnlo}) we note  that     
\begin{align}
\label{shorthands}
\mathbf{N}^T\mathbf{N} &= \left(I_{x'}+g\hat{K}_{x'}\right)^2 +
\left(I_{y'}+g\hat{K}_{y'}\right)^2  \nonumber\\
&= \mathbf{I}^2-\hat{K}_{z'}^2 +g^2\left(\hat{K}_{x'}^2+\hat{K}_{y'}^2
\right) \nonumber\\
&+2g\left(I_{x'}\hat{K}_{x'} +I_{y'}\hat{K}_{y'}\right) \ ,
\end{align}
and this expression is familiar to us from the leading-order
Hamiltonian~(\ref{Hfinal}). This makes it straight forward to evaluate 
the next-to-leading-order corrections.



%

\end{document}